\documentclass[aip,twocolumn,10pt]{revtex4-2}
\usepackage{amsmath}
\usepackage[pdftex]{graphicx}

\usepackage[usenames]{color}

\begin{document}

%\date{\today}

\author{Jan Helbing and Peter Hamm*  }

\affiliation{Department of Chemistry, University of
Zurich, Winterthurerstrasse 190, CH-8057 Zurich, Switzerland\\
*corresponding author: peter.hamm@chem.uzh.ch}

\title {\centering\large{Versatile Femtosecond Laser Synchronization for Multiple-Timescale Transient IR Spectroscopy}}

\begin{abstract}
\textbf{Abstract:} Several ways to electronically synchronize different types of amplified femtosecond laser systems are presented, based on a single freely programmable electronics hardware: Arbitrary-detuning asynchronous optical sampling, as well as actively locking two femtosecond laser oscillators, albeit not necessarily to the same round-trip frequency. They allow us to rapidly probe a very wide range of timescales, from picoseconds to potentially seconds, in a single transient absorption experiment without the need to move any delay stage. Experiments become possible that address a largely unexplored aspect of many photochemical reactions, in particular in the context of photo-catalysis as well as photoactive proteins, where an initial femtosecond trigger very often initiates a long-lasting cascade of follow-up processes. The approach is very versatile, and allows us to synchronize very different lasers, such as a Ti:Sa amplifier and a 100~kHz Yb-laser system.  The jitter of the synchronisation, and therewith the time-resolution in the transient experiment, lies in the range from 1~ps to 3~ps, depending on the method. For illustration, transient IR measurements of the excited state solvation and decay of a metal carbonyl complex as well as the full reaction cycle of bacteriorhodopsin are shown. The pros and cons of the various methods are discussed, with regard to the scientific question one might want to address, and also with regard to the laser systems that might be already existent in a laser lab.
\end{abstract}

\maketitle

\section{Introduction}
Many photochemical processes are initiated by the electronic excitation of a photo active molecule, which reacts extremely quickly on a femto- to picosecond timescale; this is the realm of femtochemistry.\cite{zewail2000} The initial process is followed by a cascade of events that may cover all timescales up to seconds or even longer. An understanding of these processes in their entirety requires a full coverage of all relevant timescales, ideally under identical conditions in a single experimental setup.

One example in this regard are photoactive proteins, in which a chromophore isomerizes after electronic excitation typically within a few 100 femtoseconds, thereby  perturbing the protein, which eventually leads to its function. For instance in the case of bacteriorhodopsin, which is probably the best studied photoactive protein, a proton is channeled via a sequence of amino acid side chains, and eventually through the membrane.\cite{oesterhelt71,siebert82,dobler88,Zscherp1997,Kuhlbrandt2000,
garczarek06,Lorenz-Fonfria2007,Nango2016} Other rhodopsins with a plethora of functions  have  been described as well,\cite{wang94,Palczewski2006} and show an equally complex cascade of responses. Other examples are photo-receptors that regulate the physiology of higher plants, microalgae, fungi and bacteria in response to environmental light conditions.\cite{Hegemann2008,Fushimi2019,Buhrke2020}, or protein conformation changes induced by an artificial photo-switch.\cite{Beharry2011,buchli13,Hull2018,Bozovic2021} %Many of these systems are considered in the context of opto-genetics.\cite{Rost2017,Deisseroth2011,Hull2018,Chernov2017}

Kinetics on multiple timescales is also encountered in photocatalytic systems that mimick natural photosynthesis. Some of their most relevant applications relate to water splitting, i.e., water reduction and oxidation,\cite{Kudo2009,Barber2009,Nocera2012,Joya2013,Favereau2016,Rodenberg15} as well as CO$_2$ reduction.\cite{Kumar2012,Sahara2015,Abdellah2017,Nitopi2019,Smieja2010,Kiefer2021} In analogy to natural photosynthesis, these systems typically separate the function of light absorption and initial charge separation in a so-called photosensitizer -- the reaction center in the natural system -- from the actual catalyst, that oxidizes/reduces the substrate. Several electron relays connect different components of the overall system. Since the latter are bimolecular reaction steps of species at low concentration, they are slow, out to the seconds timescale.\cite{Rodenberg15} While the ultrafast photophysics of the initially excited compound in these molecular systems is typically well understood from ``conventional'' femtosecond pump-probe experiments, studies of the complete reaction cycles are extremely scarce due to the lack of appropriate experimental setups that would be able to investigate them in their entirety under identical conditions and with constant sensitivity.

\begin{figure}[t]
\includegraphics[width=.45\textwidth]{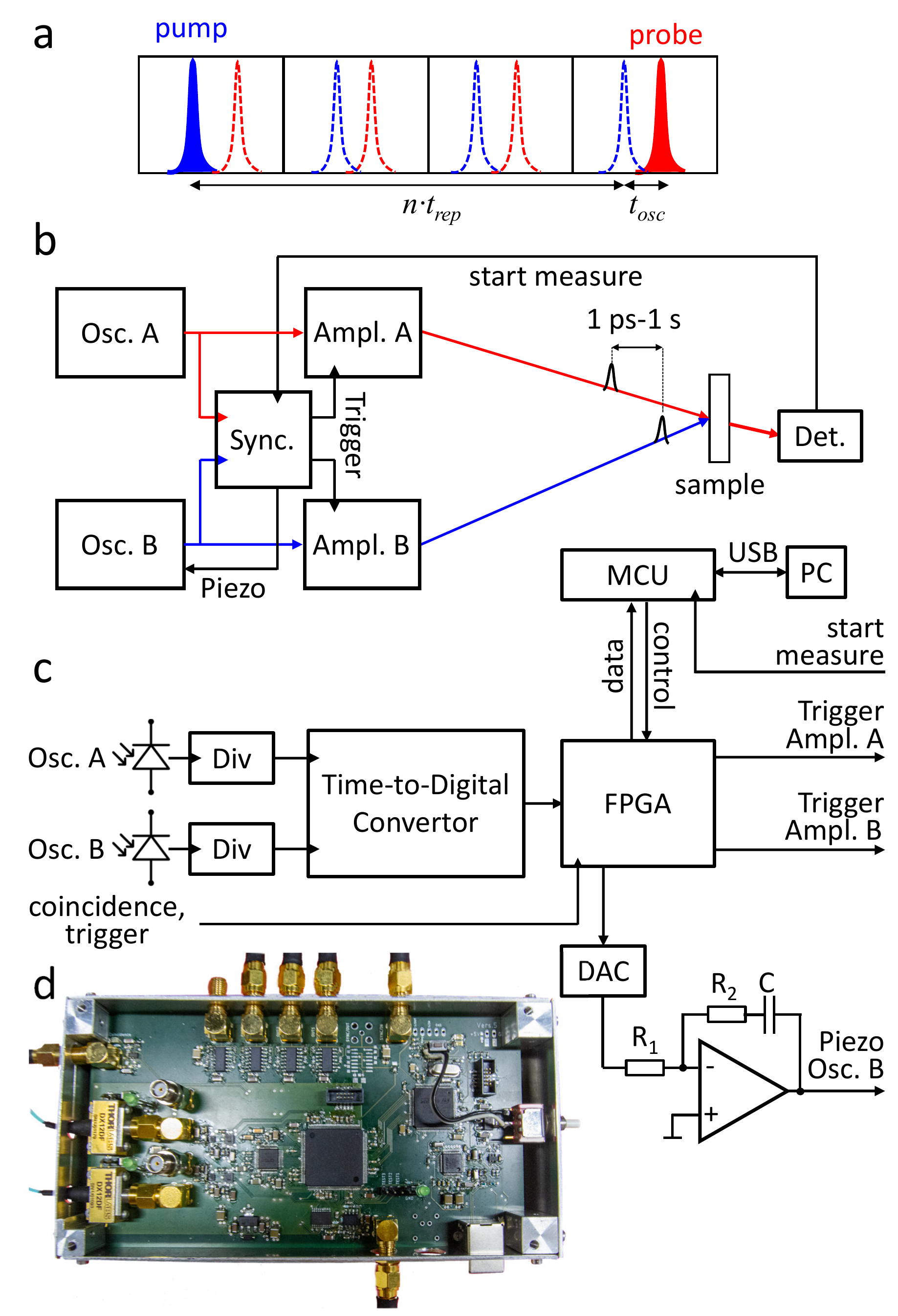}
\caption{(a) Generation of an arbitrarily short or long pump and probe delay time, that is composed of an integer-multiple of the oscillator round-trip time, $n\cdot t_{rep}$, plus the delay time between the two laser oscillators $t_{osc}$. (b) Overall laser setup, showing optical pathways of pump and probe pulses in blue and red, respectively, and electrical ones in black. (c) Schematics of the synchronisation electronic, which is explained in details in the text, and (d) picture of a prototype. } \label{FigHardware}
\end{figure}

Quite some time ago, we presented an approach that can achieve that task by electronically synchronizing two femtosecond laser systems,\cite{Bredenbeck2004} a concept that has been taken over by a few other groups.\cite{yu05b,VanWilderen2022} To understand the idea, one needs to recap that femtosecond laser systems used for these experiments consist of two components: a laser oscillator, which delivers a high-repetition frequency train of femtosecond pulses (typically many 10's of MHz, determined by the roundtrip time of the oscillator), and a regenerative amplifier, which picks individual of these pulses at a much lower repetition rate and amplifies their energy by many orders of magnitudes. In Ref.~\onlinecite{Bredenbeck2004}, the resonator length of one femtosecond Titanium:Sapphire (Ti:Sa) oscillators could be tuned with the help of a piezo actuator until its round-trip frequency locked to that of the other one. Furthermore, the relative phase of the oscillator pulse trains could be adjusted, and thereby the delay time $t_{osc}$ (see Fig.~\ref{FigHardware}a). For pump-probe delays longer than the round-trip time $t_{rep}$ of the Ti:Sa oscillators, the two Ti:Sa regenerative amplifiers were triggered in an appropriate manner to pick pulses separated by an additional integer-multiple of the oscillator repetition time, $n\cdot t_{rep}$. Similar solutions have been proposed to synchronize femtosecond lasers to a synchrotron or X-ray free-electron lasers.\cite{Xin2018}

Alternatively, one may seed two amplifiers from one common oscillator, which however requires a long (12.5~ns $\triangleq \approx$ 4~m) mechanical delay line to scan $t_{osc}$.\cite{Konold2020,Song2019} This approach has high demands on alignment and the mechanical stability of the delay line and introduces a significant dead time in the measurement, since moving a translation stage over long distances is slow. A user facility at the Rutherford Appleton Lab is probably the most advanced instrument in this regard, capable of covering timescales from $\approx100~$fs to seconds.\cite{Greetham2012,Greetham2016,Hammarback2021,Laptenok2018,Koyama2017}

A very interesting alternative is arbitrary-detuning asynchronous optical sampling (ADASOPS).\cite{Bartels2007,Antonucci2012,Antonucci2015,Solinas2017,Antonucci2020,Floery2023} The separation $t_{osc}$ between pulses from two independent oscillators is continuously changing, but can be measured, and pairs of pump and probe pulses with the desired time-delay can be selected for amplification. Particularly interesting in this regard is Ref.~\onlinecite{Antonucci2020}, which showed how to measure the time-delay purely electronically, thereby minimizing the complexity of the optical part of the setup. That is, two standard amplified femtosecond laser systems can be used without any modification. The concept builds on the fact that inexpensive time-to-digital converter chips are nowadays available, which have been developed for LiDAR applications, and which can measure the arrival times of pulses with a $\approx$10~ps single-shot accuracy (rmsd.) at a measurement rate of 10~MHz. Femtosecond laser oscillators are intrinsically very stable and do not drift on a sub-millisecond timescale. Hence, when averaging over 10000 single-shot measurements (i.e., the period between pulse selection for a kHz amplifier), an accuracy of the time measurement in the range of 100~fs can be achieved,\cite{Antonucci2020} sufficient for most chemical applications in time resolved spectroscopy.

Here, we present and compare several synchronization modes in our labs that allow us to cover a wide range of timescales: ADASOPS (Mode A, see Sec.~\ref{secSyncMode1}),  synchronisation of two identical Ti:Sa lasers systems (Mode B, see Sec.~\ref{secSyncMode2}), as well as synchronisation of a 100~kHz Yb-laser system to a Ti:Sa lasers system (Mode C, see Sec.~\ref{secSyncMode2}). The latter is particularly interesting for single-pump-multiple-probe experiments, making use of the high repetition rate of Yb-laser system.\cite{Greetham2016} All synchronization modes can be realized with one universal electronics hardware (Sec.~\ref{secHardware}). The pros and cons of the various synchronization modes will be discussed.

\section{Methods}

\subsection*{Laser Setup and Electronics Hardware} \label{secHardware}
Fig.~\ref{FigHardware}b shows the overall laser setup with two amplified femtosecond laser systems A and B, each consisting of a laser oscillator and a laser amplifier. The cavity length of Oscillator B (in our case a Tsunami/Spectra Physics) can be adjusted by roughly 1\% with a motor-driven translation stage (PD1, Thorlabs) placed inside the mount of one end-mirror. In addition, one of the resonator turning mirrors is mounted on a piezo actuator (HPSt 150, Piezomechanik, driven by a KPZ101 piezo amplifier from Thorlabs) for fine adjustment of the round-trip time (the latter is not needed for the ADASOPS Mode A).

The central synchronization electronics is shown as schematics in Fig.~\ref{FigHardware}c, and a picture of a prototype in Fig.~\ref{FigHardware}d. It is conceptually the same as that of Ref.~\onlinecite{Antonucci2020}, but quite different in its realisation. The pulse trains from the two laser oscillators are fed in by optical fibers, with fast photodiodes directly mounted on the printed circuit board (PCB) to minimize electric noise. We tested two different diodes, an ultrafast GaAs diode with a specified pulse response of 29~ps (DX12DF, Thorlabs), and a much cheaper Si PIN diode with a rise time of 700~ps (FDSP625, Thorlabs). Optionally, the signal can also be fed in via a 50~$\Omega$ SMA cable, which is required for the laser system used in Mode C.
The input signals are AC coupled into a 280~ps high-speed comparator (MAX40026, Analog Devices). The amplitude noise of laser oscillators is quite small, hence, as long as the comparator switches at a constant level, the rise time of the diode (and probably also that of the comparator) is not very critical. Indeed, we did not find any significant difference in overall performance for the two tested diodes.

The time-to-digital converter chip (AS6501, ScioSense) has three clock inputs: one reference clock as well as two stop channels, whose timings are measured relative to the reference clock. Initially, we used a separate 10~MHz crystal oscillator for the reference clock. However, laser oscillators are very stable oscillators too, and we found that the overall design becomes much simpler, and the noise performance actually better, when using one laser oscillator (oscillator B in our case) as reference clock, and measure the second laser oscillator A relative to it. The maximum reference clock frequency the time-to-digital converter can handle is $\approx$10~MHz, hence the signal of the oscillator has to be divided down accordingly, i.e. 1:8 for a 80 MHz Ti:Sa Tsunami or 1:4 for the Amplitude Tangerine SP system, whose front end seed oscillator runs at 41 MHz. To that end, we use low-jitter (1 ps) programable clock dividers (SY89873L, Microchip Technology).  The specified single-shot measurement error of the chip is 10~ps (rmsd.). Applying one and the same femtoseond laser oscillator to both inputs, we determined a single-shot noise of $\approx$12~ps, slightly more than the specification of the chip, but that also includes the jitter of the diodes, the comparators and the clock dividers. In order to improve signal-to-noise, the two stop channels of the time-to-digital converter chip are used for two independent measurements per cycle of the reference clock.

The measured data are transferred to a field-programmable gate array (FPGA, LMXO2-7000, Lattice Semiconductor Corporation) via two fast serial interfaces. The two stop events per reference clock cycle are averaged in a first step, revealing one data point with 17~bit resolution every 100~ns, related to a bit resolution of ca. 0.76~ps. The time-critical aspects of the subsequent data evaluation are also done within the FPGA according to the desired synchronisation mode. The FPGA eventually produces trigger signals for the laser amplifiers, and transfers results of the data evaluation to a microcontroller (STM32F427, STMicroelectronics), which in turn transfers them to a PC via an USB interface (FT232H, FTDI). In addition, for the  synchronisation modes B and C, the FPGA provides a feedback signal for a piezo actuator via a 12~bit, 10~MHz digital-to-analog converter (DAC, AD5445, Analog Devices). The integrator needed to close the feedback loop is realised in analog electronics, to avoid any problems due to the finite bit resolution of the DAC. The noise at the output of the DAC is larger than its bit resolution, hence is averaged out by the integrator. We did not attempt to program the integrator digitally in the FPGA (which would have the advantage that one could software-control its parameters).

One further, freely programable input of the FPGA is used in Mode A for the coincidence signal, in Mode B to synchronize the overall laser system to another device such as a photo-elastic modulator or, in Mode C for the 100~kHz signal from the Amplitude Tangerine SP system. The MCU has an additional input ``start measure'' to be able to relate the transient absorption data recorded by the detector to the corresponding delay times.

The hardware is assembled on a small $\approx$150*90~mm$^2$ PCB (see Fig.~\ref{FigHardware}d).  The FPGA is programmed in Verilog with the help of the Lattice Diamond Software (freely available from Lattice Semiconductor Corporation), and the MCU in C within Eclipse (freeware).

%\section{Synchronisation Modes} \label{secSyncMode}

\subsection*{Synchronisation Mode A: Arbitrary-Detuning Asynchronous Optical Sampling} \label{secSyncMode1}

\begin{figure}[t]
\includegraphics[width=.45\textwidth]{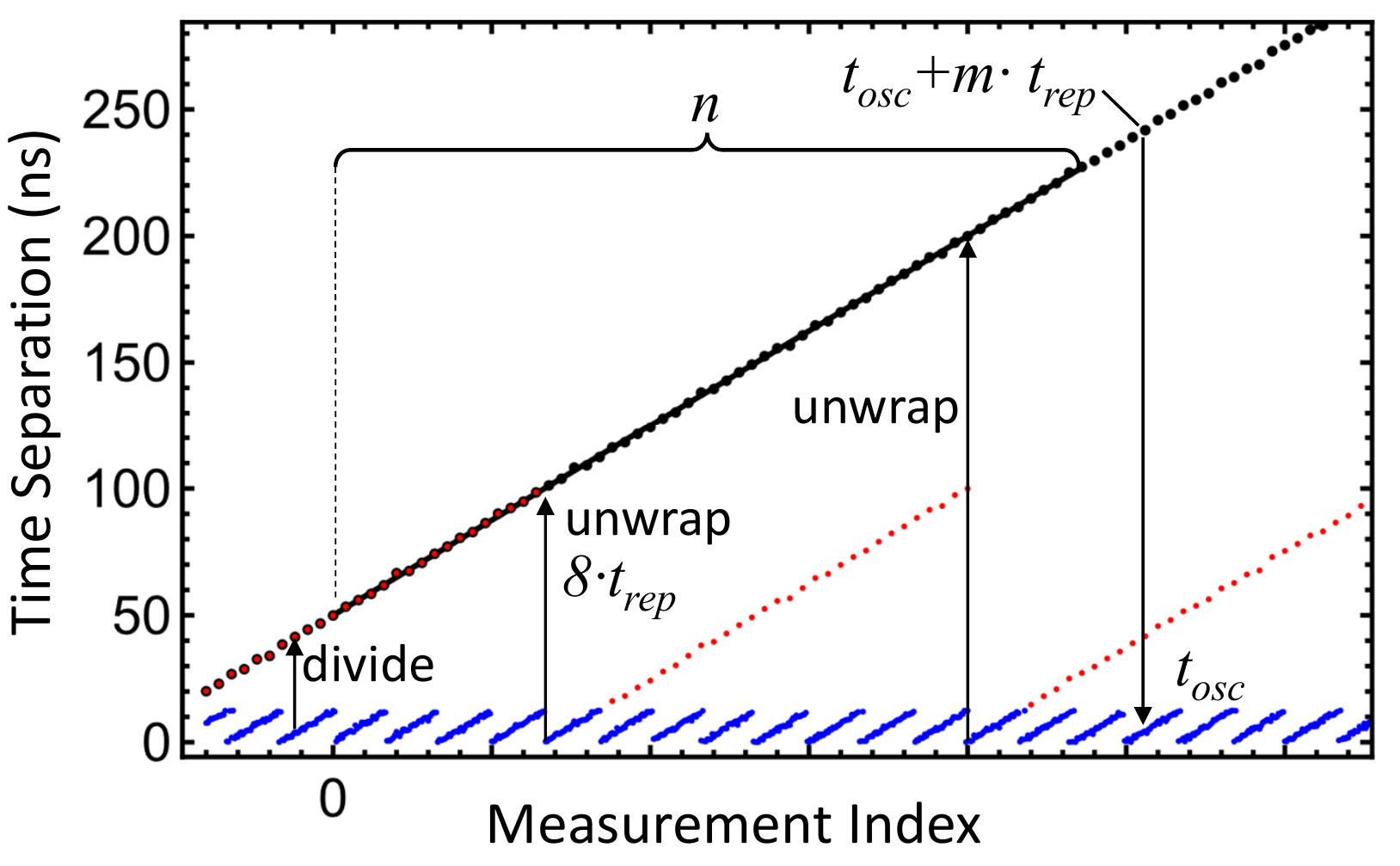}
\caption{Principle of ADASOPS. The data directly from the oscillator are shown in blue, the actually measured ones after division by a factor 8 in red, and those after unwrapping in black. Unwrapping is performed by adding eight times the oscillator round-trip time $8\cdot t_{rep}$. In total $n$ measurement points are used to perform a linear fit (black line). The fit can be extrapolated to predict the time-separation $t_{osc}+m\cdot t_{rep}$ of a later measurement point with regard to the unwrapped data, that corresponds to a time-separation between the two oscillators of $t_{osc}$. } \label{figADASOPS}
\end{figure}

The concept ADASOPS based on electronic measurements of time-separations has been introduced in Ref.~\onlinecite{Antonucci2020}, and its basic idea is illustrated in Fig.~\ref{figADASOPS}. The time-separation between the pulses from the two oscillators is shown as blue dots. In this example, oscillator A (probe laser) has a lower repetition frequency than oscillator B, which is used as a reference clock. Hence, the time of oscillator A relative to that of oscillator B increases with laser pulses. Whenever oscillator B overtakes oscillator A, the  time-separation jumps back to 0. However, as explained in Sec.~\ref{secHardware}, the time-to-digital converter chip cannot handle the repetition rate of the oscillator directly, hence the pulse train is divided down by a certain factor (8 in the example of Fig.~\ref{figADASOPS}) to reveal the dots shown in red. The slope is the same but fewer pulse are measured, the maximum time-separation becomes larger, and the rate of overtaking events is correspondingly smaller. In  addition, the data need to be unwrapped whenever an overtaking event happens for the red data points, by adding the time period of the reference clock ($8\cdot t_{rep}$), a calculation step that is performed in the FPGA.

\begin{figure*}[t]
\includegraphics[width=.9\textwidth]{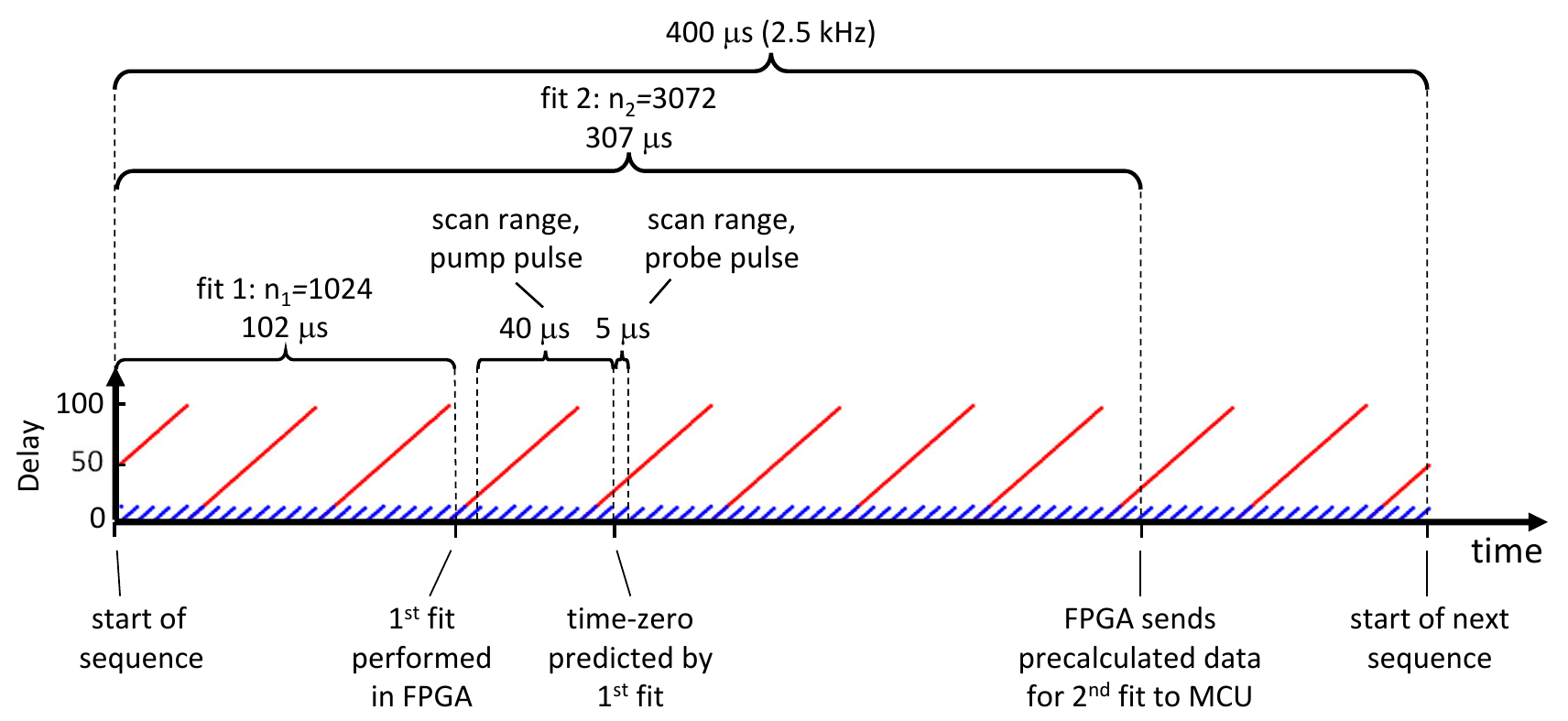}
\caption{Overview of the timing of the complete cycle between pairwise two pulses from the Ti:Sa amplifiers. The blue dots sketch the time-separations between the two Ti:Sa oscillators before division of the clock signal, and the red ones after division by a factor of 8. Important events in the cycle are marked, and are discussed in the text.} \label{figTiming}
\end{figure*}

A certain number $n$ of measurement points is fit to a model function in order to average out the noise in the single-shot data. We implemented a linear fit only, since Ref.~\onlinecite{Antonucci2020} has shown that higher-order fits, which might account for drifts in the oscillator frequencies, yield only a minor improvement of the final accuracy, and only if time-delays longer than the one anticipated here is used (i.e., 400~$\mu$s corresponding to the 2.5~kHz repetition rate of the Ti:Sa amplifier). That is, we fit the unwrapped data $t_i$ to:
\begin{equation}
  t_i=\widehat{t_0}+\widehat{\Delta t}\times i \label{eqfit}
\end{equation}
where $i=0,...,n-1$ indexes the measurement, $n$ is the number of measurements that go into the fit, $\widehat{t_0}$ the starting value for the first measurement point $i=0$, and the slope $\widehat{\Delta t}$ the change of time-separation per measurement point.

Once the fit parameters are obtained, a pulse $j>n$ can be selected by extrapolation that minimizes the distance to a desired delay $t_{osc}$. The corresponding time $t_j$ will typically include an integer-multiple of $t_{rep}$ due to dividing and  unwrapping of the raw data, see Fig.~\ref{figADASOPS}. Its index is calculated as:
\begin{equation}
   j=\frac{t_{osc}+m\cdot t_{rep}-\widehat{t_0}}{\widehat{\Delta t}}, \label{eqfit4}
\end{equation}
which is a relatively small, rounded integer number. For example, when the repetition frequencies of two 80~MHz oscillators differ by 200~kHz, the round-off error in Eq.~\ref{eqfit4} corresponds to a time-window of $\approx\pm 15$~ps, i.e., it is possible to select oscillator pulse pairs for amplification within that time window around a desired delay time $t_{osc}$. The size of the uncertainty interval is inversely proportional to the difference frequency, i.e., it could be adjusted if needed.
In any case, while this interval is relatively large, the actual time separation  for a pulse (whose index may be larger or smaller than $n$, see fit strategies below) can be obtained by evaluating Eq.~\ref{eqfit}, and this can be done  with much higher accuracy and bit resolution.

Fig.~\ref{figTiming} shows an overview of the timing of the complete cycle between two pulses from the Ti:Sa amplifiers. Our Ti:Sa amplifiers are optimized for a repetition rate of 2.5~kHz and they must be triggered by the synchronization electronic at a rate that is close to this value.  We generate the 2.5~kHz trigger signal by dividing the difference frequency between the two Ti:Sa oscillators by an integer value, i.e  200~kHz (a rather arbitrary choice) divided by 80.
This results in 10 delay-sweeps per cycle (see red points in Fig.~\ref{figTiming}). The next cycle is started when the measured delay exceeds 50~ns. Despite this being a noisy, single-shot measurement, the approach greatly helped to stabilize the overall measurement cycle by avoiding that a measurement is started at a discontinuous overtaking event.

We actually perform the linear fit described above twice, one for the first $n_1$=1024 data points, and a second one for $n_2$=3072 data points. The reasoning for this procedure is the following: The first fit is extrapolated to predict a laser shot roughly 150~$\mu$s after the beginning of the cycle with a time-separation $t_{osc}$ according to Eq.~\ref{eqfit4}, which is as close as possible to a desired value within the uncertainty interval of $\approx\pm 15$~ps. The second fit allows to determine  the actual time-separation after the fact with the help of Eq.~\ref{eqfit} with higher accuracy, since it is based on $n_2$=3072 data points (i.e. a potential factor $\sqrt{3}$ improvement), and since the selected laser shot lies in the middle of the measurement interval, and not outside (another factor $\approx$2 improvement).

\begin{figure}[t]
\includegraphics[width=.4\textwidth]{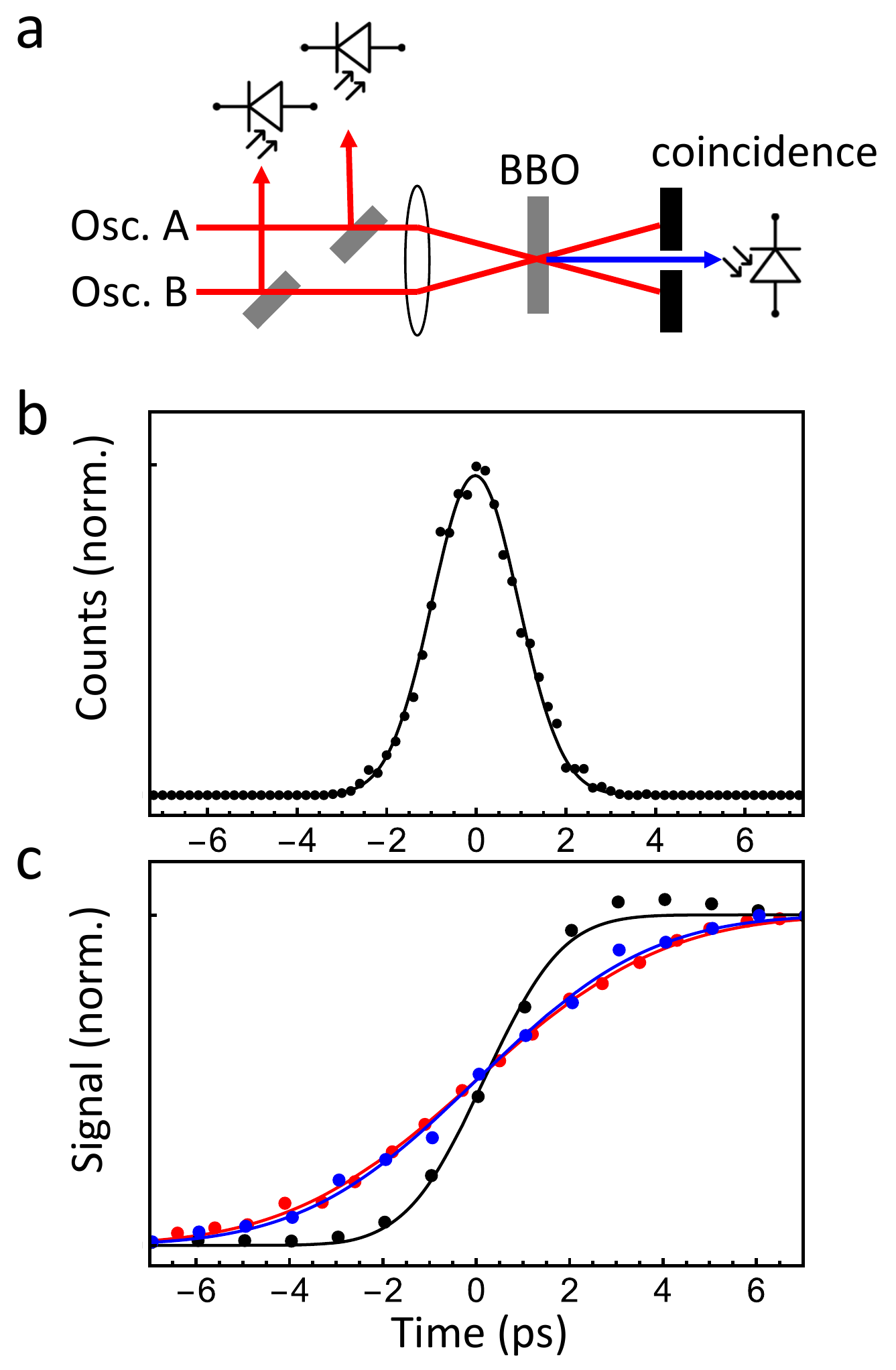}
\caption{(a) Setup for coincidence measurement between two oscillators and (b) measurement results. The data are fit to a Gaussian function with a standard deviation of 1~ps. (c) Pump-probe response of a thin Si-wafer,  showing results for ADASOPS (synchronisation Mode A) in black, those when synchronizing two identical Ti:Sa laser systems (synchronisation Mode B) in blue, and those when synchronizing a Ti:Sa laser system to a 100 kHz Yb laser system (synchronisation Mode C) in red. The data are fit to integrated Gaussain functions with standard deviations 1.3~ps, 2.9~ps, and 3.2 ps, respectively.    } \label{figCoincidence}
\end{figure}

Fig.~\ref{figCoincidence}a shows an experimental setup for a coincidence measurement, overlapping the two laser oscillators in a BBO crystal. The delay time is set to $t_{osc}=0$ within the uncertainty interval of $\approx\pm 15$~ps discussed above. A certain fraction of pulses will nevertheless coincide within $\lesssim100$~fs, i.e., the duration of the pulses, and thus generate a sum frequency signal that is detected and sent to the FPGA. Only for these coincidence events, the fitted delay times $t_j$ are collected in a histogram shown in Fig.~\ref{figCoincidence}b as dots. A fit to that histogram with a Gaussian function reveals a standard deviation of 1.0~ps, which is considered a measure of the accuracy of the time-separation measurement. The accuracy is close, but not quite as good as that reported in Ref.~\onlinecite{Antonucci2020}, for reasons that are not entirely clear. We assume that vibration isolation of the laser table becomes very critical, and our setup is not particularly optimized in this regard. That is, both laser Ti:Sa systems were on separate laser tables, which were screwed to each other.

In a pump-probe experiment, we measure transient absorption data for a sequence of time points. Despite the fact that a set delay is realized only roughly within an uncertainty interval of $\pm15$~ps, we can bin the transient absorption data according to the delay time that is actually measured with much higher precision. To that end, the electronics that reads the detector data needs to provide a signal to the synchronisation electronics to indicate the start of a measurement (see Fig.~\ref{FigHardware}b), in order to be able to relate both measurements to each other.

Fig.~\ref{figCoincidence}c, black, shows the pump-probe response of a thin Si-wafer obtained in this way, where free carriers are generated by excitation with amplified and frequency-doubled laser pulses at 420~nm. The resulting change in transmission was probed by IR pulses at $\approx$2000~cm$^{-1}$. This experiment now also includes the Ti:Sa amplifiers as well as an IR-OPA for the generation of Vis pump and IR probe pulses. ADASOPS defines delay-zero within the time-to-digital chip, and obviously, both laser pulses take very different pathways until they reach the sample. Hence, as common in femtosecond spectroscopy, there will be an arbitrary delay-zero offset, that can be large (many nanoseconds) and that has to be corrected for. In practice this offest is simply added to the (list of) delays sent from the measurement software on the PC to the MCU.

Fitting the experimental data of Fig.~\ref{figCoincidence}c, black, to an integrated Gaussian function reveals a standard deviation of 1.3~ps, slightly longer than that of the coincidence measurement of Fig.~\ref{figCoincidence}b. The data of  Figs.~\ref{figCoincidence}b and \ref{figCoincidence}c have not been measured at the same day, and we do not know whether the slight difference reflects daily performance, or the additional jitter/drifts introduced by the Ti:Sa amplifiers and the IR-OPA.

In contrast to any other method, ADASOPS does not require any mechanical motion. Hence, the delay time can be changed in a shot-by-shot manner by applying a new delay time value for the evaluations in each measurement cycle shown in Fig.~\ref{figTiming}.\cite{Solinas2017} Fast delay scanning can improve the quality of transient absorption data enormously, as it makes use of the relatively long correlation times in the noise of typical femtosecond laser systems.

However, when changing the pump-probe delay time by one full round trip time of the Ti:Sa oscillator ($\approx$12.5~ns), the position of the probe pulse in the cycle of  Fig.~\ref{figTiming} moves by~$\approx 5\mu$s. As a result, the probe amplifier will be triggered 5~$\mu$s earlier, i.e. already 395~$\mu$s after the previous shot instead of 400~$\mu$s (corresponding to the 2.5~kHz repetition rate of our Ti:Sa laser system).  The excited state lifetime of the Nd:YLF laser used to pump the Ti:Sa amplifier is 500~$\mu$s,\cite{Harmer1969} hence the excited state is not saturated after 400~$\mu$s, and a jitter in the separation between amplified pulses will be translated into energy fluctuations.  For the 1~kHz Ti:Sa laser system used in Ref.~\onlinecite{Solinas2017}, this might be much less of an issue, as the laser active medium in the Nd:YLF pump laser is close to saturation after 1~ms.

\begin{figure}[t]
\includegraphics[width=.45\textwidth]{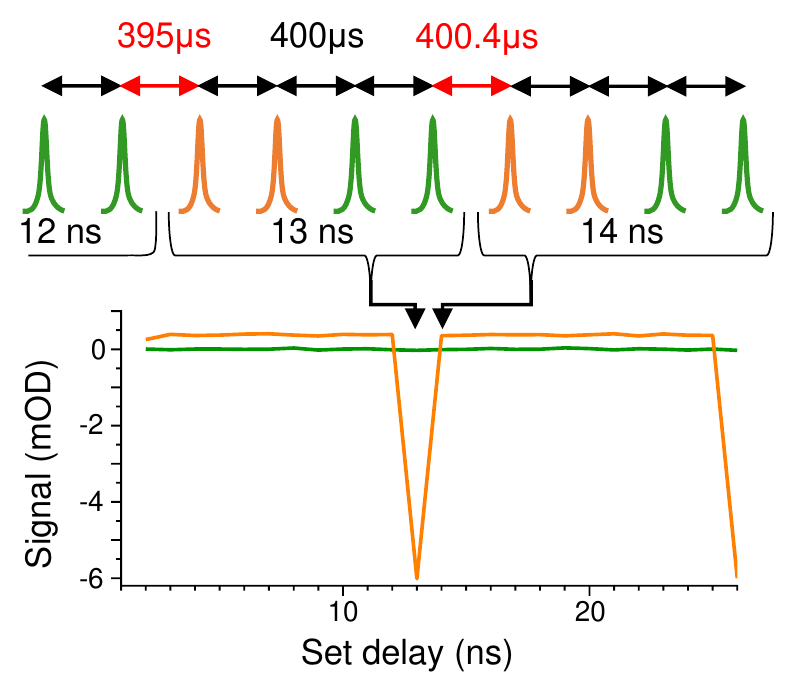}
\caption{Scanning artifacts in ADASOPS. Top: time-separation of consecutive amplified probe pulses (top arrows) when the pump-probe delay is increased by 1 ns every four laser shots. Bottom: Baseline (transient IR signal with pump beam always blocked, averaged over all detector pixels) calculated from the first and second pulses after a delay change (orange) and from the third and fourth pulses (green). } \label{figArtifact}
\end{figure}

We found that this effect can cause systematic artifacts. Fig.~\ref{figArtifact} shows the baseline (averaged over the full spectrum) recorded for a pump-probe scan where the delay was increased by 1~ns every four amplifier triggers, but the pump beam was blocked. This baseline is zero (green line) only when the signal is calculated from the third and fourth shots after every delay change (corresponding to pump on and pump off in a regular measurement). When first and second shots are used, the orange line is obtained, with a significant offset (equivalent to a signal of $\approx$ 0.5 mOD) for every ns step, and a 12 times larger 'signal' of opposite sign whenever a full oscillator round trip time is reached.

These artifacts are huge, despite the fact that we use a reference beam to normalize for probe-intensity fluctuations. The many non-linear steps needed for the mid-IR generation probably aggravate the problem. Even larger intensity changes can be expected for the amplified pump-pulses when we change the pump-probe delay by many microseconds (see Fig.~\ref{figTiming}). This is, however, less critical, because pump-pulse fluctuations only have a small effect on the already small absorption changes.
In any case, within two cycles after a delay change, the  energy of both pump and probe pulses stabilizes.

In our ADASOPS measurements, we therefore scan delay in small blocks of typically 6 amplified pulses and disregard the first two laser shots in each block. Pairwise two laser shots are chopped mechanically for referencing probe pulses with pump on and off. The experiment shown further below (Fig.~\ref{figExp}a) contains 71 delay time setting, resulting in $\approx$6 full delay-time scans per second, which is sufficient to suppress, for example, thermal laser drifts.

\subsection*{ Synchronisation Mode B: Two Identical Ti:Sa Laser Systems} \label{secSyncMode2}

This synchronisation mode is conceptually the same as our early approach,\cite{Bredenbeck2004} now realized with state-of-the-art electronics. That is, the time-separations between the pulses of the two Ti:Sa oscillators is measured in a digital manner, as described for synchronisation Mode A, which produces an error-signal that is fed back to a piezo-actuator in one of the Ti:Sa oscillators. It changes the resonator length of this Ti:Sa oscillator, and hence its frequency, until both oscillators are locked. An additional number can be added to the error signal in the FPGA before outputting it to the DAC, hence, the two oscillators lock with a software-defined time-separation $t_{osc}$ between them.

In the language of electronics, synchronizing two Ti:Sa oscillators is a phase-locked loop (PLL), where the piezo-controlled Ti:Sa oscillator acts as the voltage controlled oscillator (VCO), whose frequency changes with the voltage applied to the piezo-actuator. The theory of PLL's is well established, see e.g. Ref.~\onlinecite{Best2007}.
%and is briefly repeated in Appendix~\ref{AppendixLoop} to facilitate setting the feedback loop parameters.
We achieved good performance when setting the loop natural frequency to 20~Hz with close to critical damping (these values can be adjusted with R$_1$, R$_2$ and C in the integrator shown in Fig.~\ref{FigHardware}c). Increasing the loop natural frequency beyond 20~Hz resulted in a $\approx$300~Hz oscillation of the feedback loop, presumably due to additional time constants introduced into the overall loop filter due to the slow response of the piezo-actuator.

Fig.~\ref{figCoincidence}c, blue, shows the pump-probe response of a Si-wafer measured with two synchronized Ti:Sa laser systems. The fit reveals a jitter with a standard deviation of 2.9~ps, about a factor two larger that that of ADASOPS (Fig.~\ref{figCoincidence}c, black).

\subsection*{Synchronisation Mode C: Ti:Sa Laser System Synchronized to a 100~kHz Yb Laser System} \label{secSyncMode3}

High-repetition rate Yb-based femtosecond laser systems become more common in transient IR spectroscopy spectroscopy.\cite{Donaldson2023}   This new technology not only promises better signal-to-noise ratios due to the higher repetition rates and better stability, but also enables conceptually new experiments. This is in particular single-pump-multiple probe transient absorption spectroscopy, which may cover an extremely wide range of timescales from picoseconds to second with one and the same experimental setup.\cite{Greetham2016}

To realize such an instrument, we synchronize a Tangerin SP laser system (Amplitude) to a Ti:Sa laser system (Tsunami/Spitfire, Spectra Physics). The former produces a train of probe pulses separated by 10~$\mu$s each, the latter delivers pump pulses at a lower repetition rate, ranging from 1~Hz to 2.5~kHz, which can be adapted to the length of the reaction of the investigated sample. Within the first 10~$\mu$s, the relative timing of the  two lasers systems needs to be scanned, analogous to the synchronisation methods described above. Beyond 10~$\mu$s, we get transient absorption data ``for free'', since the Yb-laser produces additional probe pulses.

ADASOPS will not work in this case. Setting the delay time to a desired value requires that we have a certain freedom to chose which oscillator pulse is being amplified in both pump and probe lasers. Practically speaking, the Tangerin SP system does not accept any trigger input. But even if that were  possible, the jitter this may introduce could produce unacceptably high pulse energy fluctuations, an additional problem beyond that discussed with Fig.~\ref{figArtifact}: When both lasers run asynchronously, a jitter of at least one oscillator pulse would necessarily be introduced, even if the desired delay time is kept constant. This is no longer negligible relative to the repetition rate of the amplifier.

\begin{figure*}[t]
\includegraphics[width=.9\textwidth]{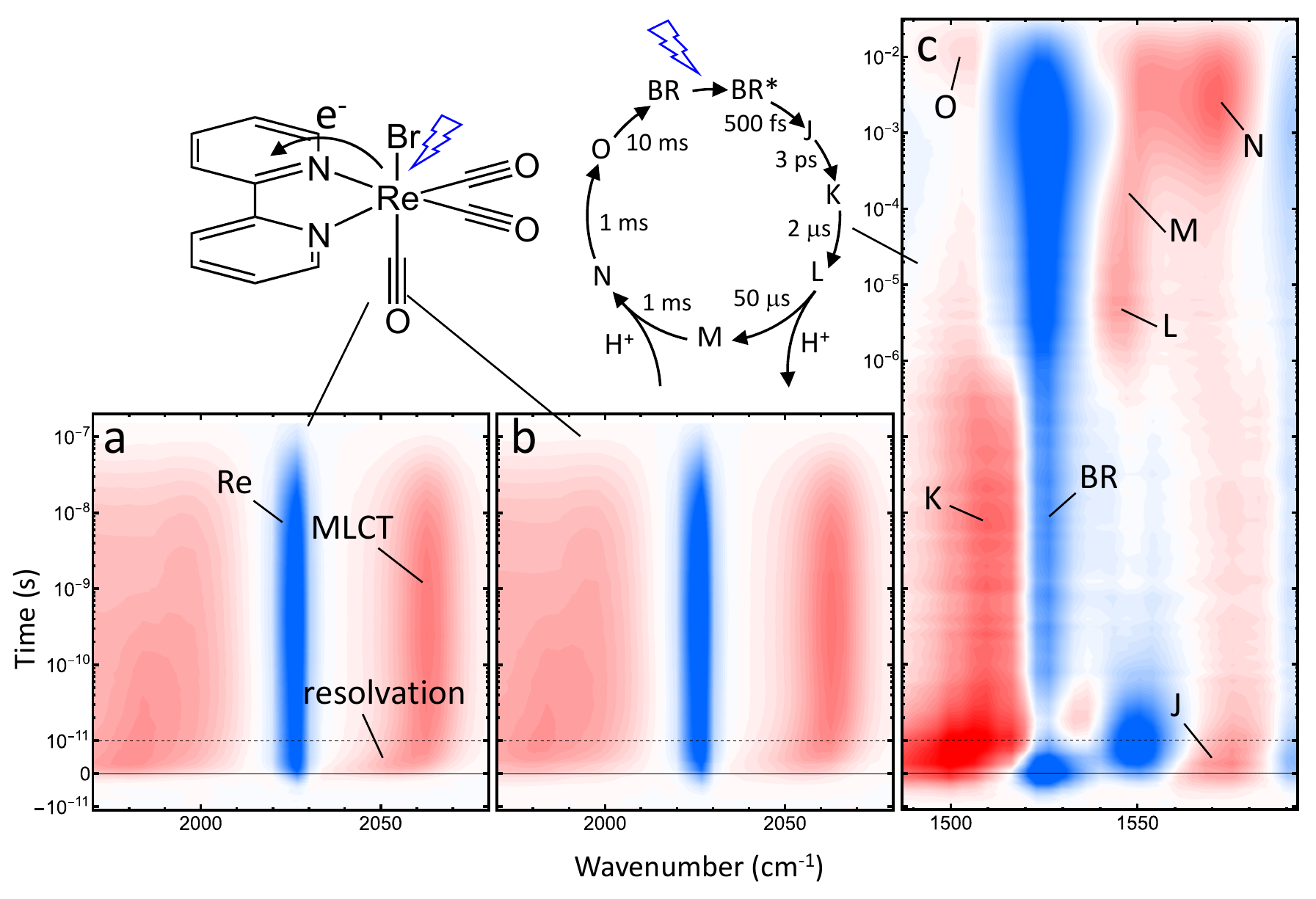}
\caption{Demonstration experiments for the three different synchronisation modes: in (a) with ADASOPS (synchronisation Modes A), in (b)  two identical and synchronized Ti:Sa laser systems (synchronisation Modes B), and (c) with a Ti:Sa laser system synchronized to a 100 kHz Yb laser system (synchronisation Modes C). In panels (a) and (b), the photo-induced MLCT and excitetd state decay of Re(bpy)(CO)$_3$Br in DMSO is measured, in panel (c) the photocycle of bacteriorhodopsin in D$_2$O. The time range from -10~ps to +10~ps is plotted on a linear scale, beyond on a logarithmic scale. Negative bleaches are shown in blue, positive transient bands in red. In the top-left corner, the MLCT in Re(bpy)(CO)$_3$Br is shown, as well as the photo-cycle of bacteriorhodopsin including approximate times of the reaction steps. Marker modes of the various intermediates are indicated in the transient spectra (see text for details).} \label{figExp}
\end{figure*}

Synchronisation Mode B will also not work, at least not directly, since the oscillator frequencies are too far apart. That is, the Yb-oscillator runs at 41~MHz, and the Ti:Sa oscillator at $\approx80.4$~MHz, the latter tunable by $\approx\pm0.2$~MHz with the help of the additional motor-driven mirror mount. However, both lasers do not need to be locked for each individual laser pulse, they need to be locked only every 10~$\mu$s, when a pulse from the Yb-oscillator is  picked by the Yb-amplifier. That is, the frequencies of the Yb-oscillator and the Ti:Sa oscillator can be off by an integer-multiple of 100~kHz, which is within the tuning range of the Ti:Sa oscillator. Hence, the synchronisation mode described here is a mix of synchronisation Modes A and B, i.e., we synchronize two asynchronous laser oscillators, locking them to a defined difference frequency.

To that end, we calculate in the FPGA the unwrapped data shown in Fig.~\ref{figADASOPS}, subtract from them a linear function with properly chosen slope, and output the result as an error signal to the DAC. The integrator and piezo-actuator in the Ti:Sa oscillator close the feedback loop just as in synchronisation Modes B. The same can be achieved much more precisely - albeit with severe constraints on the laser system and repetition rates, with dual-comb lasers, which have very recently been used produce amplified fs-pulses at arbitrary delay\cite{Floery2023}.

Fig.~\ref{figCoincidence}c, red, shows the pump-probe response of the Si-wafer measured in this configuration. The standard deviation of 3.2~ps is about the same as in synchronisation mode B. Here it is important to note that both laser systems were in different (neighboring) rooms on disconnected laser tables, with the Ti:Sa pump pulse guided through a hole in the wall over $\approx$4~m from one laser table to the other. The oscillator signal needed for the synchronisation was sent via an optical fibre. Somewhat surprisingly, we did not see any drift of time-zero over a period of $\approx$1~h. If drifts  became an issue, one could send the oscillator beam in parallel to the amplified beam from one laser table to the other, and pick it up in a fibre only then. In that way, one could compensate for distance drifts between two laser tables.

\section{Results} \label{secDemo}

Fig.~\ref{figExp} shows demonstration experiments for the three different synchronisation modes. In all of these experiments, the sample was excited with the 2nd harmonic of a Ti:Sa laser system (Tsunami/Spitfire, Spectra Physics). The probe pulses were derived from a home-built mid-IR-OPA pumped with an identical Ti:Sa laser system system in Fig.~\ref{figExp}a,b,\cite{ham00b} and  a Twin STARZZ (Fastlite) pumped with a Tangerine SP (Amplitude) in Fig.~\ref{figExp}c.
For referencing, in all examples the probe pulses were split into two beams, only one of which in overlap with the pumped volume of the sample, and imaged onto one line each of a two-line MCT detector with single-shot readout.\cite{Farrell2020}
To compute absorption changes, in modes A and B the pump pulses were mechanically chopped at half the repetition frequency of the pump amplifier. In mode C several probe pulses were acquired before the arrival of each pump pulse.

Figs.~\ref{figExp}a,b show the response of the C=O modes of Re(bpy)CO$_3$Br to photo-induced metal-to-ligand charge transfer (MLCT), with the focus on the symmetric stretch vibration. Upon photo-excitation, an electron is transferred to the ligand, formally oxidizing the Re-center, which causes a characteristic blue shift of the CO vibrations due to $\pi$-backbonding. Initially, the blueshift is small, and then increases on a 4~ps timescale due to the resolvation of the new charge distribution in the charge transfer state.\cite{asbury02} That process is resolved by both methods, i.e., ADASOPS (Fig.~\ref{figExp}a) as well as with the two synchronized Ti:Sa laser systems (Fig.~\ref{figExp}b), but is more distinct in the former case due to the higher time-resolution. The lifetime of the $^3$MLCT state is about 50~ns, hence the plots of Figs.~\ref{figExp} show data up to 150~ns, when the signal is fully decayed.

Fig.~\ref{figExp}c explores the full reaction cycle of bacteriorhodopsin.\cite{oesterhelt71,Kuhlbrandt2000} Bacteriorhodopsin is a light driven proton pump consisting of an $\alpha$-helical \textit{trans}-membrane protein with a retinal molecule bound to the protein via a Schiff-base. Photo-excitation induces the \textit{trans}-to-\textit{cis} isomerization of the chomophore around one of its C=C double bonds on a 500~fs timescale.\cite{dobler88} It then proceeds through a sequence of intermediates commonly denoted as J to O. The Schiff-base is deprotonated during the L-M transition and the proton is transferred an Asp side chain, and is reprotonated from another Asp during the M-N transition. Thermal back-isomerisation of the chromophore occurs during the N-O transition, and deprotonation of the Asp sidechain completes the catalytic cycle.\cite{Kuhlbrandt2000} Most transient IR studies have been performed with ``slower'' spectroscopic techniques, such as step-scan FTIR spectroscopy\cite{siebert82,uhmann91,Gerwert1993,Zscherp1997,garczarek06,Lorenz-Fonfria2007} or more recently with the help of quantum cascade lasers,\cite{Kottke2017,Klocke2018,Stritt2020,Schubert2022} while the fs-ps timescale has been addressed separately.\cite{herbst02,Gross2009a}

Synchronisation Mode C allows us to cover all timescales at once, see Fig.~\ref{figExp}c. The dominant feature in the transient absorption spectra is the negative (blue) bleach of an ethylenic stretch vibration of the retinal chromophore (labeled BR in Fig.~\ref{figExp}c). This band exhibits the strongest response, since it is affected directly by its photo-isomerization.\cite{smith87} The band shifts to different frequencies in the various states of the photo cycle, showing up as positive (red) bands in Fig.~\ref{figExp}c. The time resolution of the experiment, $\approx$3~ps (see Fig.~\ref{figCoincidence}c, red), is sufficient to resolve the earliest J state, whose lifetime is $\approx$3~ps.\cite{Polland1986,herbst02} All subsequent states, i.e., state K through state O, reveal characteristic marker modes that are indicated in Fig.~\ref{figExp}c. The final state O relaxes back to the resting state BR on a 10~ms timescale. Consequently, the repetition frequency of the pump pulses has been set to $\approx$30~Hz in this experiment to assure that subsequent pump pulses hit a fully relaxed sample. To do this without affecting laser performance, we only adjusted the Pockels cell triggers of the Ti:S amplifier, but did not change the 2.5 kHz repetition rate of its YLF pump laser. The sample was not exchanged on a shot-by-shot basis, but we found that it bleaches on a minutes timescale, hence the sample cuvette was slowly rasterized.

\begin{table*}
\caption{Summary of features of the various synchronisation modes} \label{TabFeatures}
\begin{tabular}{l|lll}
\hline
\hline
& \shortstack{ Mode A \\ ADASOPS} &  \shortstack{ Mode B\\pump-probe}& \shortstack{ Mode C\\single-pump-multiple probe}\\
\hline
Probe rep. rate $f_{probe}^{amp}$ & $\approx$ 2.5 kHz & 2.5 kHz & 100 kHz \\
Modification of pump oscillator & none & motor+piezo & motor+piezo\\
Pump oscillator frequency $f_{pump}^{osc}$ & freely running\hspace{1cm}  & locked to $f_{probe}^{osc}$\hspace{0.5cm} &  locked to $f_{probe}^{osc}+k f_{probe}^{amp}$    \\
At set delay &  $\approx$ every 5~$\mu$s & always &  every 10~$\mu$s  \\
Time resolution & 1.3~ps & 2.9~ps & 3.2~ps \\
Scan range & $\approx40~\mu$s & $\approx40~\mu$s & $>1$~s \\
Fast scanning & yes & no & for delays $>10~\mu$s\\
\hline
\end{tabular}
\end{table*}

\section{Discussion and Conclusion}
Table~\ref{TabFeatures} compares some of  the features of the various synchronisation modes. They are complementary to a certain extent and have their pros and cons. For example, since the absolute timing is variable within certain limits, but does not change when changing the delay times in synchronisation Mode B, the two amplifiers can be synchronized to yet another device, such as a photo-elastic modulator (PEM) used for polarisation control with its own fixed resonance frequency.\cite{Helbing09}

On the other hand, ADASOPS (synchronisation Mode A) can work without any modification of the laser systems. We derive the repetition frequency of the Ti:Sa amplifiers from the difference frequency of the two oscillators by division through a certain factor. The exact repetition frequency is not critical (as long as it stays constant) and could be adjusted for any given laser system by changing that division factor. Having a motor-driven translation stage in one of the oscillators to adjust its round-trip time is convenient, but not required, and neither is the piezo actuator. We anticipate that this capability will be the most important application of ADASOPS. Furthermore, since ADASOPS involves no moving parts, not even the micrometer motion of a piezo actuator, it enables fast (i.e., shot-by-shot) scanning of the pump-probe delay time (with the limitation discussed in Sec.~\ref{secSyncMode3}), which is advantageous in terms of signal-to-noise of transient data. Finally, the time resolution is better with ADASOPS.

Unfortunately, ADASOPS is not easily compatible with Yb-laser systems, as it requires a certain freedom with regard to selecting the oscillator pulse that is sent to the amplifier; the high repetition rate of these lasers is rather restrictive in this regard. We see the biggest potential of  high-repetition rate Yb-laser systems in single-pump-multiple probe experiments that allow one to cover all chemical relevant timescales,\cite{Greetham2016} ranging from the initial photochemical trigger of the photoactive part of a molecular system until it reaches its final state, which often spans a time range from picoseconds to seconds. Synchronisation Mode C enables such experiments.

The focus in our lab lies on transient IR spectroscopy, which is sensitive to chemical structure. However, since nonlinear optical processes can be driven with high yields due to the high peak power of amplified femtosecond laser systems, other spectroscopic regimes, ranging from THz to X-ray, are equally accessible. At least for transient IR spectroscopy, the somewhat reduced time resolution of 1-3~ps is not limiting, as perturbed-free induction decay typically limits the response time of vibrational transitions to that time regime.\cite{Hamm1995,Wynne1995}

One important prerequisite that makes these concepts possible is modern FPGAs, which render hardware software-programmable. The style of programming of FPGAs is quite different from ``regular'' programming languages, but is learnable for anybody with programming experience (some experience in ``conventional'' digital electronics helps as well). To learn how to program FPGAs, we recommend to start from a small development board, such as tinyFPGA.

In our current realisation, we implemented synchronisation Modes A and B into one FPGA with a software switch to select them without the need to re-plug any cable in the laser setup. Switching between the two synchronisation modes is thus a matter of minutes. Mode C, which we use in a different laser lab, is currently implemented in another FPGA, yet utilizing the same electronics hardware. One can consider Mode B a special case of  Mode C with difference frequency 0, and implementing all 3 modes in a single FPGA would certainly be possible. In any case, the universal implementation of the three methods allows us to freely select the most suitable method for a certain problem. In fact, all three data sets shown in Fig.~\ref{figExp} were acquired on a single day. The flexibility of the FPGAs make it possible to adapt the synchronisation electronics to other laser configurations and/or other synchronisation modes with only a limited programming effort.

To conclude, we consider the synchronisation electronics shown in Fig.~\ref{FigHardware}d a very versatile tool for all sorts of timing and synchronisation tasks of femtosecond laser systems. In particular, many femtosecond labs that acquire a new high-repetition rate laser system can probably recycle the old Ti:Sa laser it replaces; the older the better since a non-integrated construct, such as the Tsunami oscillator from Spectra Physics, makes it easier to add the motor-driven translation stage and piezo-actuator needed to tune its round-trip time. With rather limited effort, they may combine old and new laser systems into a (still) quite unique spectrometer that can address a largely unexplored time-window of photochemical reactions.\\

\noindent\textbf{Conflict of Interest:} The authors have no conflicts of interest to disclose.\\

%\R{\noindent\textbf{Disclaimer:} For reasons of scientific completeness, we need to specify  the Yb-laser system used in this study, which is not meant to be a recommendation for the particular system.\\}

\noindent\textbf{Data Availability Statement:}
The data used in this article are available on Zenodo (https://doi.org/10.5281/zenodo.8085588).
Upon a reasonable request, hardware and software of the synchronisation electronics will be made available in the framework of a collaboration.\\

\noindent\textbf{Acknowledgement:}  We thank Kerstin Oppelt and David Buhrke for providing Re(bpy)CO$_3$Br and BR samples, respectively. The work has been supported by the Swiss National Science Foundation (SNF) through Grants 200020B\_188694/1 and 200020\_192240.\\

%ACS style of citation, i.e., puts citations in brackets
%\makeatletter
%\def\@biblabel#1{(#1)}
%\makeatother

%\bibliographystyle{achemso}
%\setkeys{acs}{maxauthors = 0,articletitle=true}

%\bibliography{../../../library}

\begin{thebibliography}{66}%
\makeatletter
\providecommand \@ifxundefined [1]{%
 \@ifx{#1\undefined}
}%
\providecommand \@ifnum [1]{%
 \ifnum #1\expandafter \@firstoftwo
 \else \expandafter \@secondoftwo
 \fi
}%
\providecommand \@ifx [1]{%
 \ifx #1\expandafter \@firstoftwo
 \else \expandafter \@secondoftwo
 \fi
}%
\providecommand \natexlab [1]{#1}%
\providecommand \enquote  [1]{``#1''}%
\providecommand \bibnamefont  [1]{#1}%
\providecommand \bibfnamefont [1]{#1}%
\providecommand \citenamefont [1]{#1}%
\providecommand \href@noop [0]{\@secondoftwo}%
\providecommand \href [0]{\begingroup \@sanitize@url \@href}%
\providecommand \@href[1]{\@@startlink{#1}\@@href}%
\providecommand \@@href[1]{\endgroup#1\@@endlink}%
\providecommand \@sanitize@url [0]{\catcode `\\12\catcode `\$12\catcode
  `\&12\catcode `\#12\catcode `\^12\catcode `\_12\catcode `\%12\relax}%
\providecommand \@@startlink[1]{}%
\providecommand \@@endlink[0]{}%
\providecommand \url  [0]{\begingroup\@sanitize@url \@url }%
\providecommand \@url [1]{\endgroup\@href {#1}{\urlprefix }}%
\providecommand \urlprefix  [0]{URL }%
\providecommand \Eprint [0]{\href }%
\providecommand \doibase [0]{https://doi.org/}%
\providecommand \selectlanguage [0]{\@gobble}%
\providecommand \bibinfo  [0]{\@secondoftwo}%
\providecommand \bibfield  [0]{\@secondoftwo}%
\providecommand \translation [1]{[#1]}%
\providecommand \BibitemOpen [0]{}%
\providecommand \bibitemStop [0]{}%
\providecommand \bibitemNoStop [0]{.\EOS\space}%
\providecommand \EOS [0]{\spacefactor3000\relax}%
\providecommand \BibitemShut  [1]{\csname bibitem#1\endcsname}%
\let\auto@bib@innerbib\@empty
%</preamble>
\bibitem [{\citenamefont {Zewail}(2000)}]{zewail2000}%
  \BibitemOpen
  \bibfield  {author} {\bibinfo {author} {\bibfnamefont {A.~H.}\ \bibnamefont
  {Zewail}},\ }\bibfield  {title} {\enquote {\bibinfo {title} {{Femtochemistry:
  Atomic-Scale Dynamics of the Chemical Bond}},}\ }\href@noop {} {\bibfield
  {journal} {\bibinfo  {journal} {J. Phys. Chem. A.}\ }\textbf {\bibinfo
  {volume} {104}},\ \bibinfo {pages} {5660--5694} (\bibinfo {year}
  {2000})}\BibitemShut {NoStop}%
\bibitem [{\citenamefont {Oesterhelt}\ and\ \citenamefont
  {Stoeckenius}(1971)}]{oesterhelt71}%
  \BibitemOpen
  \bibfield  {author} {\bibinfo {author} {\bibfnamefont {D.}~\bibnamefont
  {Oesterhelt}}\ and\ \bibinfo {author} {\bibfnamefont {W.}~\bibnamefont
  {Stoeckenius}},\ }\bibfield  {title} {\enquote {\bibinfo {title}
  {{Rhodopsin-like protein from the purple membrane of Halobacterium
  halobium}},}\ }\href@noop {} {\bibfield  {journal} {\bibinfo  {journal}
  {Nature, New Biol.}\ }\textbf {\bibinfo {volume} {233}},\ \bibinfo {pages}
  {149--152} (\bibinfo {year} {1971})}\BibitemShut {NoStop}%
\bibitem [{\citenamefont {Siebert}, \citenamefont {M{\"{a}}ntele},\ and\
  \citenamefont {Kreutz}(1982)}]{siebert82}%
  \BibitemOpen
  \bibfield  {author} {\bibinfo {author} {\bibfnamefont {F.}~\bibnamefont
  {Siebert}}, \bibinfo {author} {\bibfnamefont {W.}~\bibnamefont
  {M{\"{a}}ntele}},\ and\ \bibinfo {author} {\bibfnamefont {W.}~\bibnamefont
  {Kreutz}},\ }\bibfield  {title} {\enquote {\bibinfo {title} {{Evidence for
  the protonation of two internal carboxylic groups during the photocycle of
  bacteriorhodopsin Investigation of kinetic infrared spectroscopy}},}\
  }\href@noop {} {\bibfield  {journal} {\bibinfo  {journal} {Febs Lett.}\
  }\textbf {\bibinfo {volume} {141}},\ \bibinfo {pages} {82--87} (\bibinfo
  {year} {1982})}\BibitemShut {NoStop}%
\bibitem [{\citenamefont {Dobler}\ \emph {et~al.}(1988)\citenamefont {Dobler},
  \citenamefont {Zinth}, \citenamefont {Kaiser},\ and\ \citenamefont
  {Oesterhelt}}]{dobler88}%
  \BibitemOpen
  \bibfield  {author} {\bibinfo {author} {\bibfnamefont {J.}~\bibnamefont
  {Dobler}}, \bibinfo {author} {\bibfnamefont {W.}~\bibnamefont {Zinth}},
  \bibinfo {author} {\bibfnamefont {W.}~\bibnamefont {Kaiser}},\ and\ \bibinfo
  {author} {\bibfnamefont {D.}~\bibnamefont {Oesterhelt}},\ }\bibfield  {title}
  {\enquote {\bibinfo {title} {{Excited-State Reaction Dynamics of
  Bacteriorhodopsin Studied by Femtosecond Spectroscopy.}}}\ }\href@noop {}
  {\bibfield  {journal} {\bibinfo  {journal} {Chem. Phys. Lett.}\ }\textbf
  {\bibinfo {volume} {144}},\ \bibinfo {pages} {215} (\bibinfo {year}
  {1988})}\BibitemShut {NoStop}%
\bibitem [{\citenamefont {Zscherp}\ and\ \citenamefont
  {Heberle}(1997)}]{Zscherp1997}%
  \BibitemOpen
  \bibfield  {author} {\bibinfo {author} {\bibfnamefont {C.}~\bibnamefont
  {Zscherp}}\ and\ \bibinfo {author} {\bibfnamefont {J.}~\bibnamefont
  {Heberle}},\ }\bibfield  {title} {\enquote {\bibinfo {title} {{Infrared
  difference spectra of the intermediates L, M, N, and O of the
  bacteriorhodopsin photoreaction obtained by time-resolved attenuated total
  reflection spectroscopy}},}\ }\href {https://doi.org/10.1021/jp971047i}
  {\bibfield  {journal} {\bibinfo  {journal} {J. Phys. Chem. B}\ }\textbf
  {\bibinfo {volume} {101}},\ \bibinfo {pages} {10542--10547} (\bibinfo {year}
  {1997})}\BibitemShut {NoStop}%
\bibitem [{\citenamefont {K{\"{u}}hlbrandt}(2000)}]{Kuhlbrandt2000}%
  \BibitemOpen
  \bibfield  {author} {\bibinfo {author} {\bibfnamefont {W.}~\bibnamefont
  {K{\"{u}}hlbrandt}},\ }\bibfield  {title} {\enquote {\bibinfo {title}
  {{Bacteriorhodopsin - the movie.}}}\ }\href
  {https://doi.org/10.1038/35020654} {\bibfield  {journal} {\bibinfo  {journal}
  {Nature}\ }\textbf {\bibinfo {volume} {406}},\ \bibinfo {pages} {569--570}
  (\bibinfo {year} {2000})}\BibitemShut {NoStop}%
\bibitem [{\citenamefont {Garczarek}\ and\ \citenamefont
  {Gerwert}(2006)}]{garczarek06}%
  \BibitemOpen
  \bibfield  {author} {\bibinfo {author} {\bibfnamefont {F.}~\bibnamefont
  {Garczarek}}\ and\ \bibinfo {author} {\bibfnamefont {K.}~\bibnamefont
  {Gerwert}},\ }\bibfield  {title} {\enquote {\bibinfo {title} {{Functional
  waters in intraprotein proton transfer monitored by FTIR difference
  spectroscopy}},}\ }\href@noop {} {\bibfield  {journal} {\bibinfo  {journal}
  {Nature}\ }\textbf {\bibinfo {volume} {439}},\ \bibinfo {pages} {109--112}
  (\bibinfo {year} {2006})}\BibitemShut {NoStop}%
\bibitem [{\citenamefont {L{\'{o}}renz-Fonfr{\'{i}}a}\ and\ \citenamefont
  {Kandori}(2007)}]{Lorenz-Fonfria2007}%
  \BibitemOpen
  \bibfield  {author} {\bibinfo {author} {\bibfnamefont {V.~A.}\ \bibnamefont
  {L{\'{o}}renz-Fonfr{\'{i}}a}}\ and\ \bibinfo {author} {\bibfnamefont
  {H.}~\bibnamefont {Kandori}},\ }\bibfield  {title} {\enquote {\bibinfo
  {title} {{Bayesian maximum entropy (two-dimensional) lifetime distribution
  reconstruction from time-resolved spectroscopic data}},}\ }\href
  {https://doi.org/10.1366/000370207780466172} {\bibfield  {journal} {\bibinfo
  {journal} {Appl. Spectrosc.}\ }\textbf {\bibinfo {volume} {61}},\ \bibinfo
  {pages} {428--443} (\bibinfo {year} {2007})}\BibitemShut {NoStop}%
\bibitem [{\citenamefont {Nango}\ \emph {et~al.}(2016)\citenamefont {Nango},
  \citenamefont {Royant}, \citenamefont {Kubo}, \citenamefont {Nakane},
  \citenamefont {Wickstrand}, \citenamefont {Kimura}, \citenamefont {Tanaka},
  \citenamefont {Tono}, \citenamefont {Song}, \citenamefont {Tanaka},
  \citenamefont {Arima}, \citenamefont {Yamashita}, \citenamefont {Kobayashi},
  \citenamefont {Hosaka}, \citenamefont {Mizohata}, \citenamefont {Nogly},
  \citenamefont {Sugahara}, \citenamefont {Nam}, \citenamefont {Nomura},
  \citenamefont {Shimamura}, \citenamefont {Im}, \citenamefont {Fujiwara},
  \citenamefont {Yamanaka}, \citenamefont {Jeon}, \citenamefont {Nishizawa},
  \citenamefont {Oda}, \citenamefont {Fukuda}, \citenamefont {Andersson},
  \citenamefont {B{\aa}th}, \citenamefont {Dods}, \citenamefont {Davidsson},
  \citenamefont {Matsuoka}, \citenamefont {Kawatake}, \citenamefont {Murata},
  \citenamefont {Nureki}, \citenamefont {Owada}, \citenamefont {Kameshima},
  \citenamefont {Hatsui}, \citenamefont {Joti}, \citenamefont {Schertler},
  \citenamefont {Yabashi}, \citenamefont {Bondar}, \citenamefont {Standfuss},
  \citenamefont {Neutze},\ and\ \citenamefont {Iwata}}]{Nango2016}%
  \BibitemOpen
  \bibfield  {author} {\bibinfo {author} {\bibfnamefont {E.}~\bibnamefont
  {Nango}}, \bibinfo {author} {\bibfnamefont {A.}~\bibnamefont {Royant}},
  \bibinfo {author} {\bibfnamefont {M.}~\bibnamefont {Kubo}}, \bibinfo {author}
  {\bibfnamefont {T.}~\bibnamefont {Nakane}}, \bibinfo {author} {\bibfnamefont
  {C.}~\bibnamefont {Wickstrand}}, \bibinfo {author} {\bibfnamefont
  {T.}~\bibnamefont {Kimura}}, \bibinfo {author} {\bibfnamefont
  {T.}~\bibnamefont {Tanaka}}, \bibinfo {author} {\bibfnamefont
  {K.}~\bibnamefont {Tono}}, \bibinfo {author} {\bibfnamefont {C.}~\bibnamefont
  {Song}}, \bibinfo {author} {\bibfnamefont {R.}~\bibnamefont {Tanaka}},
  \bibinfo {author} {\bibfnamefont {T.}~\bibnamefont {Arima}}, \bibinfo
  {author} {\bibfnamefont {A.}~\bibnamefont {Yamashita}}, \bibinfo {author}
  {\bibfnamefont {J.}~\bibnamefont {Kobayashi}}, \bibinfo {author}
  {\bibfnamefont {T.}~\bibnamefont {Hosaka}}, \bibinfo {author} {\bibfnamefont
  {E.}~\bibnamefont {Mizohata}}, \bibinfo {author} {\bibfnamefont
  {P.}~\bibnamefont {Nogly}}, \bibinfo {author} {\bibfnamefont
  {M.}~\bibnamefont {Sugahara}}, \bibinfo {author} {\bibfnamefont
  {D.}~\bibnamefont {Nam}}, \bibinfo {author} {\bibfnamefont {T.}~\bibnamefont
  {Nomura}}, \bibinfo {author} {\bibfnamefont {T.}~\bibnamefont {Shimamura}},
  \bibinfo {author} {\bibfnamefont {D.}~\bibnamefont {Im}}, \bibinfo {author}
  {\bibfnamefont {T.}~\bibnamefont {Fujiwara}}, \bibinfo {author}
  {\bibfnamefont {Y.}~\bibnamefont {Yamanaka}}, \bibinfo {author}
  {\bibfnamefont {B.}~\bibnamefont {Jeon}}, \bibinfo {author} {\bibfnamefont
  {T.}~\bibnamefont {Nishizawa}}, \bibinfo {author} {\bibfnamefont
  {K.}~\bibnamefont {Oda}}, \bibinfo {author} {\bibfnamefont {M.}~\bibnamefont
  {Fukuda}}, \bibinfo {author} {\bibfnamefont {R.}~\bibnamefont {Andersson}},
  \bibinfo {author} {\bibfnamefont {P.}~\bibnamefont {B{\aa}th}}, \bibinfo
  {author} {\bibfnamefont {R.}~\bibnamefont {Dods}}, \bibinfo {author}
  {\bibfnamefont {J.}~\bibnamefont {Davidsson}}, \bibinfo {author}
  {\bibfnamefont {S.}~\bibnamefont {Matsuoka}}, \bibinfo {author}
  {\bibfnamefont {S.}~\bibnamefont {Kawatake}}, \bibinfo {author}
  {\bibfnamefont {M.}~\bibnamefont {Murata}}, \bibinfo {author} {\bibfnamefont
  {O.}~\bibnamefont {Nureki}}, \bibinfo {author} {\bibfnamefont
  {S.}~\bibnamefont {Owada}}, \bibinfo {author} {\bibfnamefont
  {T.}~\bibnamefont {Kameshima}}, \bibinfo {author} {\bibfnamefont
  {T.}~\bibnamefont {Hatsui}}, \bibinfo {author} {\bibfnamefont
  {Y.}~\bibnamefont {Joti}}, \bibinfo {author} {\bibfnamefont {G.}~\bibnamefont
  {Schertler}}, \bibinfo {author} {\bibfnamefont {M.}~\bibnamefont {Yabashi}},
  \bibinfo {author} {\bibfnamefont {A.-N.}\ \bibnamefont {Bondar}}, \bibinfo
  {author} {\bibfnamefont {J.}~\bibnamefont {Standfuss}}, \bibinfo {author}
  {\bibfnamefont {R.}~\bibnamefont {Neutze}},\ and\ \bibinfo {author}
  {\bibfnamefont {S.}~\bibnamefont {Iwata}},\ }\bibfield  {title} {\enquote
  {\bibinfo {title} {{A three-dimensional movie of structural changes in
  bacteriorhodopsin}},}\ }\href {https://doi.org/10.1126/science.aah3497}
  {\bibfield  {journal} {\bibinfo  {journal} {Science}\ }\textbf {\bibinfo
  {volume} {354}},\ \bibinfo {pages} {1552--1557} (\bibinfo {year}
  {2016})}\BibitemShut {NoStop}%
\bibitem [{\citenamefont {Wang}\ \emph {et~al.}(1994)\citenamefont {Wang},
  \citenamefont {Schoenlein}, \citenamefont {Peteanu}, \citenamefont
  {Mathies},\ and\ \citenamefont {Shank}}]{wang94}%
  \BibitemOpen
  \bibfield  {author} {\bibinfo {author} {\bibfnamefont {Q.}~\bibnamefont
  {Wang}}, \bibinfo {author} {\bibfnamefont {R.~W.}\ \bibnamefont
  {Schoenlein}}, \bibinfo {author} {\bibfnamefont {L.~A.}\ \bibnamefont
  {Peteanu}}, \bibinfo {author} {\bibfnamefont {R.~A.}\ \bibnamefont
  {Mathies}},\ and\ \bibinfo {author} {\bibfnamefont {C.~V.}\ \bibnamefont
  {Shank}},\ }\bibfield  {title} {\enquote {\bibinfo {title} {{Vibrationally
  coherent photochemistry in the femtosecond primary event of vision}},}\
  }\href@noop {} {\bibfield  {journal} {\bibinfo  {journal} {Science}\ }\textbf
  {\bibinfo {volume} {266}},\ \bibinfo {pages} {422--424} (\bibinfo {year}
  {1994})}\BibitemShut {NoStop}%
\bibitem [{\citenamefont {Palczewski}(2006)}]{Palczewski2006}%
  \BibitemOpen
  \bibfield  {author} {\bibinfo {author} {\bibfnamefont {K.}~\bibnamefont
  {Palczewski}},\ }\bibfield  {title} {\enquote {\bibinfo {title} {{G
  protein-coupled receptor rhodopsin}},}\ }\href
  {https://doi.org/10.1146/annurev.biochem.75.103004.142743} {\bibfield
  {journal} {\bibinfo  {journal} {Annu. Rev. Biochem.}\ }\textbf {\bibinfo
  {volume} {75}},\ \bibinfo {pages} {743--767} (\bibinfo {year}
  {2006})}\BibitemShut {NoStop}%
\bibitem [{\citenamefont {Hegemann}(2008)}]{Hegemann2008}%
  \BibitemOpen
  \bibfield  {author} {\bibinfo {author} {\bibfnamefont {P.}~\bibnamefont
  {Hegemann}},\ }\bibfield  {title} {\enquote {\bibinfo {title} {{Algal sensory
  photoreceptors}},}\ }\href
  {https://doi.org/10.1146/annurev.arplant.59.032607.092847} {\bibfield
  {journal} {\bibinfo  {journal} {Annu. Rev. Plant Biol.}\ }\textbf {\bibinfo
  {volume} {59}},\ \bibinfo {pages} {167--189} (\bibinfo {year}
  {2008})}\BibitemShut {NoStop}%
\bibitem [{\citenamefont {Fushimi}\ and\ \citenamefont
  {Narikawa}(2019)}]{Fushimi2019}%
  \BibitemOpen
  \bibfield  {author} {\bibinfo {author} {\bibfnamefont {K.}~\bibnamefont
  {Fushimi}}\ and\ \bibinfo {author} {\bibfnamefont {R.}~\bibnamefont
  {Narikawa}},\ }\bibfield  {title} {\enquote {\bibinfo {title}
  {{Cyanobacteriochromes: photoreceptors covering the entire UV-to-visible
  spectrum}},}\ }\href {https://doi.org/10.1016/j.sbi.2019.01.018} {\bibfield
  {journal} {\bibinfo  {journal} {Curr. Opin. Struct. Biol.}\ }\textbf
  {\bibinfo {volume} {57}},\ \bibinfo {pages} {39--46} (\bibinfo {year}
  {2019})}\BibitemShut {NoStop}%
\bibitem [{\citenamefont {Buhrke}\ \emph {et~al.}(2020)\citenamefont {Buhrke},
  \citenamefont {Oppelt}, \citenamefont {Heckmeier}, \citenamefont
  {Fernandez-Teran},\ and\ \citenamefont {Hamm}}]{Buhrke2020}%
  \BibitemOpen
  \bibfield  {author} {\bibinfo {author} {\bibfnamefont {D.}~\bibnamefont
  {Buhrke}}, \bibinfo {author} {\bibfnamefont {K.~T.}\ \bibnamefont {Oppelt}},
  \bibinfo {author} {\bibfnamefont {P.~J.}\ \bibnamefont {Heckmeier}}, \bibinfo
  {author} {\bibfnamefont {R.}~\bibnamefont {Fernandez-Teran}},\ and\ \bibinfo
  {author} {\bibfnamefont {P.}~\bibnamefont {Hamm}},\ }\bibfield  {title}
  {\enquote {\bibinfo {title} {{Nanosecond protein dynamics in a red/green
  Cyanobacteriochrome revealed by transient IR spectroscopy}},}\ }\href
  {https://doi.org/10.1063/5.0033107} {\bibfield  {journal} {\bibinfo
  {journal} {J. Chem. Phys.}\ }\textbf {\bibinfo {volume} {153}},\ \bibinfo
  {pages} {245101} (\bibinfo {year} {2020})}\BibitemShut {NoStop}%
\bibitem [{\citenamefont {Beharry}\ and\ \citenamefont
  {Woolley}(2011)}]{Beharry2011}%
  \BibitemOpen
  \bibfield  {author} {\bibinfo {author} {\bibfnamefont {A.~A.}\ \bibnamefont
  {Beharry}}\ and\ \bibinfo {author} {\bibfnamefont {G.~A.}\ \bibnamefont
  {Woolley}},\ }\bibfield  {title} {\enquote {\bibinfo {title} {{Azobenzene
  photoswitches for biomolecules}},}\ }\href
  {https://doi.org/10.1039/c1cs15023e} {\bibfield  {journal} {\bibinfo
  {journal} {Chem. Soc. Rev.}\ }\textbf {\bibinfo {volume} {40}},\ \bibinfo
  {pages} {4422--4437} (\bibinfo {year} {2011})}\BibitemShut {NoStop}%
\bibitem [{\citenamefont {Buchli}\ \emph {et~al.}(2013)\citenamefont {Buchli},
  \citenamefont {Waldauer}, \citenamefont {Walser}, \citenamefont {Donten},
  \citenamefont {Pfister}, \citenamefont {Bl{\"{o}}chliger}, \citenamefont
  {Steiner}, \citenamefont {Caflisch}, \citenamefont {Zerbe},\ and\
  \citenamefont {Hamm}}]{buchli13}%
  \BibitemOpen
  \bibfield  {author} {\bibinfo {author} {\bibfnamefont {B.}~\bibnamefont
  {Buchli}}, \bibinfo {author} {\bibfnamefont {S.~A.}\ \bibnamefont
  {Waldauer}}, \bibinfo {author} {\bibfnamefont {R.}~\bibnamefont {Walser}},
  \bibinfo {author} {\bibfnamefont {M.~L.}\ \bibnamefont {Donten}}, \bibinfo
  {author} {\bibfnamefont {R.}~\bibnamefont {Pfister}}, \bibinfo {author}
  {\bibfnamefont {N.}~\bibnamefont {Bl{\"{o}}chliger}}, \bibinfo {author}
  {\bibfnamefont {S.}~\bibnamefont {Steiner}}, \bibinfo {author} {\bibfnamefont
  {A.}~\bibnamefont {Caflisch}}, \bibinfo {author} {\bibfnamefont
  {O.}~\bibnamefont {Zerbe}},\ and\ \bibinfo {author} {\bibfnamefont
  {P.}~\bibnamefont {Hamm}},\ }\bibfield  {title} {\enquote {\bibinfo {title}
  {{Kinetic response of a photoperturbed allosteric protein}},}\ }\href@noop {}
  {\bibfield  {journal} {\bibinfo  {journal} {Proc. Natl. Acad. Sci. USA}\
  }\textbf {\bibinfo {volume} {110}},\ \bibinfo {pages} {11725--11730}
  (\bibinfo {year} {2013})}\BibitemShut {NoStop}%
\bibitem [{\citenamefont {H{\"{u}}ll}, \citenamefont {Morstein},\ and\
  \citenamefont {Trauner}(2018)}]{Hull2018}%
  \BibitemOpen
  \bibfield  {author} {\bibinfo {author} {\bibfnamefont {K.}~\bibnamefont
  {H{\"{u}}ll}}, \bibinfo {author} {\bibfnamefont {J.}~\bibnamefont
  {Morstein}},\ and\ \bibinfo {author} {\bibfnamefont {D.}~\bibnamefont
  {Trauner}},\ }\bibfield  {title} {\enquote {\bibinfo {title} {{In Vivo
  Photopharmacology}},}\ }\href {https://doi.org/10.1021/acs.chemrev.8b00037}
  {\bibfield  {journal} {\bibinfo  {journal} {Chem. Rev.}\ }\textbf {\bibinfo
  {volume} {118}},\ \bibinfo {pages} {10710--10747} (\bibinfo {year}
  {2018})}\BibitemShut {NoStop}%
\bibitem [{\citenamefont {Bozovic}\ \emph {et~al.}(2021)\citenamefont
  {Bozovic}, \citenamefont {Ruf}, \citenamefont {Zanobini}, \citenamefont
  {Jankovic}, \citenamefont {Buhrke}, \citenamefont {Johnson},\ and\
  \citenamefont {Hamm}}]{Bozovic2021}%
  \BibitemOpen
  \bibfield  {author} {\bibinfo {author} {\bibfnamefont {O.}~\bibnamefont
  {Bozovic}}, \bibinfo {author} {\bibfnamefont {J.}~\bibnamefont {Ruf}},
  \bibinfo {author} {\bibfnamefont {C.}~\bibnamefont {Zanobini}}, \bibinfo
  {author} {\bibfnamefont {B.}~\bibnamefont {Jankovic}}, \bibinfo {author}
  {\bibfnamefont {D.}~\bibnamefont {Buhrke}}, \bibinfo {author} {\bibfnamefont
  {P.~J.~M.}\ \bibnamefont {Johnson}},\ and\ \bibinfo {author} {\bibfnamefont
  {P.}~\bibnamefont {Hamm}},\ }\bibfield  {title} {\enquote {\bibinfo {title}
  {{The Speed of Allosteric Signaling Within a Single-Domain Protein}},}\
  }\href@noop {} {\bibfield  {journal} {\bibinfo  {journal} {J. Phys. Chem.
  Lett.}\ }\textbf {\bibinfo {volume} {12}},\ \bibinfo {pages} {4262--4267}
  (\bibinfo {year} {2021})}\BibitemShut {NoStop}%
\bibitem [{\citenamefont {Kudo}\ and\ \citenamefont {Miseki}(2009)}]{Kudo2009}%
  \BibitemOpen
  \bibfield  {author} {\bibinfo {author} {\bibfnamefont {A.}~\bibnamefont
  {Kudo}}\ and\ \bibinfo {author} {\bibfnamefont {Y.}~\bibnamefont {Miseki}},\
  }\bibfield  {title} {\enquote {\bibinfo {title} {{Heterogeneous photocatalyst
  materials for water splitting}},}\ }\href {https://doi.org/10.1039/b800489g}
  {\bibfield  {journal} {\bibinfo  {journal} {Chem. Soc. Rev.}\ }\textbf
  {\bibinfo {volume} {38}},\ \bibinfo {pages} {253--278} (\bibinfo {year}
  {2009})}\BibitemShut {NoStop}%
\bibitem [{\citenamefont {Barber}(2009)}]{Barber2009}%
  \BibitemOpen
  \bibfield  {author} {\bibinfo {author} {\bibfnamefont {J.}~\bibnamefont
  {Barber}},\ }\bibfield  {title} {\enquote {\bibinfo {title} {{Photosynthetic
  energy conversion: natural and artificial.}}}\ }\href
  {https://doi.org/10.1039/b802262n} {\bibfield  {journal} {\bibinfo  {journal}
  {Chem. Soc. Rev.}\ }\textbf {\bibinfo {volume} {38}},\ \bibinfo {pages}
  {185--96} (\bibinfo {year} {2009})}\BibitemShut {NoStop}%
\bibitem [{\citenamefont {Nocera}(2012)}]{Nocera2012}%
  \BibitemOpen
  \bibfield  {author} {\bibinfo {author} {\bibfnamefont {D.~G.}\ \bibnamefont
  {Nocera}},\ }\bibfield  {title} {\enquote {\bibinfo {title} {{The artificial
  leaf}},}\ }\href {https://doi.org/10.1021/ar2003013} {\bibfield  {journal}
  {\bibinfo  {journal} {Acc. Chem. Res.}\ }\textbf {\bibinfo {volume} {45}},\
  \bibinfo {pages} {767--776} (\bibinfo {year} {2012})}\BibitemShut {NoStop}%
\bibitem [{\citenamefont {Joya}\ \emph {et~al.}(2013)\citenamefont {Joya},
  \citenamefont {Joya}, \citenamefont {Ocakoglu},\ and\ \citenamefont {{Van De
  Krol}}}]{Joya2013}%
  \BibitemOpen
  \bibfield  {author} {\bibinfo {author} {\bibfnamefont {K.~S.}\ \bibnamefont
  {Joya}}, \bibinfo {author} {\bibfnamefont {Y.~F.}\ \bibnamefont {Joya}},
  \bibinfo {author} {\bibfnamefont {K.}~\bibnamefont {Ocakoglu}},\ and\
  \bibinfo {author} {\bibfnamefont {R.}~\bibnamefont {{Van De Krol}}},\
  }\bibfield  {title} {\enquote {\bibinfo {title} {{Water-splitting catalysis
  and solar fuel devices: Artificial leaves on the move}},}\ }\href
  {https://doi.org/10.1002/anie.201300136} {\bibfield  {journal} {\bibinfo
  {journal} {Angew. Chemie - Int. Ed.}\ }\textbf {\bibinfo {volume} {52}},\
  \bibinfo {pages} {10426--10437} (\bibinfo {year} {2013})}\BibitemShut
  {NoStop}%
\bibitem [{\citenamefont {Favereau}\ \emph {et~al.}(2016)\citenamefont
  {Favereau}, \citenamefont {Makhal}, \citenamefont {Pellegrin}, \citenamefont
  {Blart}, \citenamefont {Petersson}, \citenamefont {G{\"{o}}ransson},
  \citenamefont {Hammarstr{\"{o}}m},\ and\ \citenamefont
  {Odobel}}]{Favereau2016}%
  \BibitemOpen
  \bibfield  {author} {\bibinfo {author} {\bibfnamefont {L.}~\bibnamefont
  {Favereau}}, \bibinfo {author} {\bibfnamefont {A.}~\bibnamefont {Makhal}},
  \bibinfo {author} {\bibfnamefont {Y.}~\bibnamefont {Pellegrin}}, \bibinfo
  {author} {\bibfnamefont {E.}~\bibnamefont {Blart}}, \bibinfo {author}
  {\bibfnamefont {J.}~\bibnamefont {Petersson}}, \bibinfo {author}
  {\bibfnamefont {E.}~\bibnamefont {G{\"{o}}ransson}}, \bibinfo {author}
  {\bibfnamefont {L.}~\bibnamefont {Hammarstr{\"{o}}m}},\ and\ \bibinfo
  {author} {\bibfnamefont {F.}~\bibnamefont {Odobel}},\ }\bibfield  {title}
  {\enquote {\bibinfo {title} {{A Molecular Tetrad That Generates a High-Energy
  Charge-Separated State by Mimicking the Photosynthetic Z-Scheme}},}\ }\href
  {https://doi.org/10.1021/jacs.5b12650} {\bibfield  {journal} {\bibinfo
  {journal} {J. Am. Chem. Soc.}\ }\textbf {\bibinfo {volume} {138}},\ \bibinfo
  {pages} {3752--3760} (\bibinfo {year} {2016})}\BibitemShut {NoStop}%
\bibitem [{\citenamefont {Rodenberg}\ \emph {et~al.}(2015)\citenamefont
  {Rodenberg}, \citenamefont {Orazietti}, \citenamefont {Probst}, \citenamefont
  {Bachman}, \citenamefont {Alberto}, \citenamefont {Baldridge},\ and\
  \citenamefont {Hamm}}]{Rodenberg15}%
  \BibitemOpen
  \bibfield  {author} {\bibinfo {author} {\bibfnamefont {A.}~\bibnamefont
  {Rodenberg}}, \bibinfo {author} {\bibfnamefont {M.}~\bibnamefont
  {Orazietti}}, \bibinfo {author} {\bibfnamefont {B.}~\bibnamefont {Probst}},
  \bibinfo {author} {\bibfnamefont {C.}~\bibnamefont {Bachman}}, \bibinfo
  {author} {\bibfnamefont {R.}~\bibnamefont {Alberto}}, \bibinfo {author}
  {\bibfnamefont {K.~K.}\ \bibnamefont {Baldridge}},\ and\ \bibinfo {author}
  {\bibfnamefont {P.}~\bibnamefont {Hamm}},\ }\bibfield  {title} {\enquote
  {\bibinfo {title} {{Mechanism of Photocatalytic Hydrogen Generation by a
  Polypyridyl- Based Cobalt Catalyst in Aqueous Solution}},}\ }\href@noop {}
  {\bibfield  {journal} {\bibinfo  {journal} {Inorg. Chem.}\ }\textbf {\bibinfo
  {volume} {54}},\ \bibinfo {pages} {646--657} (\bibinfo {year}
  {2015})}\BibitemShut {NoStop}%
\bibitem [{\citenamefont {Kumar}\ \emph {et~al.}(2012)\citenamefont {Kumar},
  \citenamefont {Llorente}, \citenamefont {Froehlich}, \citenamefont {Dang},
  \citenamefont {Sathrum},\ and\ \citenamefont {Kubiak}}]{Kumar2012}%
  \BibitemOpen
  \bibfield  {author} {\bibinfo {author} {\bibfnamefont {B.}~\bibnamefont
  {Kumar}}, \bibinfo {author} {\bibfnamefont {M.}~\bibnamefont {Llorente}},
  \bibinfo {author} {\bibfnamefont {J.}~\bibnamefont {Froehlich}}, \bibinfo
  {author} {\bibfnamefont {T.}~\bibnamefont {Dang}}, \bibinfo {author}
  {\bibfnamefont {A.}~\bibnamefont {Sathrum}},\ and\ \bibinfo {author}
  {\bibfnamefont {C.~P.}\ \bibnamefont {Kubiak}},\ }\bibfield  {title}
  {\enquote {\bibinfo {title} {{Photochemical and photoelectrochemical
  reduction of CO2.}}}\ }\href
  {https://doi.org/10.1146/annurev-physchem-032511-143759} {\bibfield
  {journal} {\bibinfo  {journal} {Annu. Rev. Phys. Chem.}\ }\textbf {\bibinfo
  {volume} {63}},\ \bibinfo {pages} {541--569} (\bibinfo {year}
  {2012})}\BibitemShut {NoStop}%
\bibitem [{\citenamefont {Sahara}\ and\ \citenamefont
  {Ishitani}(2015)}]{Sahara2015}%
  \BibitemOpen
  \bibfield  {author} {\bibinfo {author} {\bibfnamefont {G.}~\bibnamefont
  {Sahara}}\ and\ \bibinfo {author} {\bibfnamefont {O.}~\bibnamefont
  {Ishitani}},\ }\bibfield  {title} {\enquote {\bibinfo {title} {{Efficient
  Photocatalysts for CO2 Reduction}},}\ }\href
  {https://doi.org/10.1021/ic502675a} {\bibfield  {journal} {\bibinfo
  {journal} {Inorg. Chem.}\ }\textbf {\bibinfo {volume} {54}},\ \bibinfo
  {pages} {5096--5104} (\bibinfo {year} {2015})}\BibitemShut {NoStop}%
\bibitem [{\citenamefont {Abdellah}\ \emph {et~al.}(2017)\citenamefont
  {Abdellah}, \citenamefont {El-Zohry}, \citenamefont {Antila}, \citenamefont
  {Windle}, \citenamefont {Reisner},\ and\ \citenamefont
  {Hammarstr{\"{o}}m}}]{Abdellah2017}%
  \BibitemOpen
  \bibfield  {author} {\bibinfo {author} {\bibfnamefont {M.}~\bibnamefont
  {Abdellah}}, \bibinfo {author} {\bibfnamefont {A.~M.}\ \bibnamefont
  {El-Zohry}}, \bibinfo {author} {\bibfnamefont {L.~J.}\ \bibnamefont
  {Antila}}, \bibinfo {author} {\bibfnamefont {C.~D.}\ \bibnamefont {Windle}},
  \bibinfo {author} {\bibfnamefont {E.}~\bibnamefont {Reisner}},\ and\ \bibinfo
  {author} {\bibfnamefont {L.}~\bibnamefont {Hammarstr{\"{o}}m}},\ }\bibfield
  {title} {\enquote {\bibinfo {title} {{Time-Resolved IR Spectroscopy Reveals a
  Mechanism with TiO2 as a Reversible Electron Acceptor in a TiO2 – Re
  Catalyst CO2 Photoreduction System}},}\ }\href
  {https://doi.org/10.1021/jacs.6b11308} {\bibfield  {journal} {\bibinfo
  {journal} {J. Am. Chem. Soc.}\ }\textbf {\bibinfo {volume} {139}},\ \bibinfo
  {pages} {1226--1232} (\bibinfo {year} {2017})}\BibitemShut {NoStop}%
\bibitem [{\citenamefont {Nitopi}\ \emph {et~al.}(2019)\citenamefont {Nitopi},
  \citenamefont {Bertheussen}, \citenamefont {Scott}, \citenamefont {Liu},
  \citenamefont {Engstfeld}, \citenamefont {Horch}, \citenamefont {Seger},
  \citenamefont {Stephens}, \citenamefont {Chan}, \citenamefont {Hahn},
  \citenamefont {N{\o}rskov}, \citenamefont {Jaramillo},\ and\ \citenamefont
  {Chorkendorff}}]{Nitopi2019}%
  \BibitemOpen
  \bibfield  {author} {\bibinfo {author} {\bibfnamefont {S.}~\bibnamefont
  {Nitopi}}, \bibinfo {author} {\bibfnamefont {E.}~\bibnamefont {Bertheussen}},
  \bibinfo {author} {\bibfnamefont {S.~B.}\ \bibnamefont {Scott}}, \bibinfo
  {author} {\bibfnamefont {X.}~\bibnamefont {Liu}}, \bibinfo {author}
  {\bibfnamefont {A.~K.}\ \bibnamefont {Engstfeld}}, \bibinfo {author}
  {\bibfnamefont {S.}~\bibnamefont {Horch}}, \bibinfo {author} {\bibfnamefont
  {B.}~\bibnamefont {Seger}}, \bibinfo {author} {\bibfnamefont {I.~E.}\
  \bibnamefont {Stephens}}, \bibinfo {author} {\bibfnamefont {K.}~\bibnamefont
  {Chan}}, \bibinfo {author} {\bibfnamefont {C.}~\bibnamefont {Hahn}}, \bibinfo
  {author} {\bibfnamefont {J.~K.}\ \bibnamefont {N{\o}rskov}}, \bibinfo
  {author} {\bibfnamefont {T.~F.}\ \bibnamefont {Jaramillo}},\ and\ \bibinfo
  {author} {\bibfnamefont {I.}~\bibnamefont {Chorkendorff}},\ }\bibfield
  {title} {\enquote {\bibinfo {title} {{Progress and Perspectives of
  Electrochemical CO2 Reduction on Copper in Aqueous Electrolyte}},}\ }\href
  {https://doi.org/10.1021/acs.chemrev.8b00705} {\bibfield  {journal} {\bibinfo
   {journal} {Chem. Rev.}\ }\textbf {\bibinfo {volume} {119}},\ \bibinfo
  {pages} {7610--7672} (\bibinfo {year} {2019})}\BibitemShut {NoStop}%
\bibitem [{\citenamefont {Smieja}\ and\ \citenamefont
  {Kubiak}(2010)}]{Smieja2010}%
  \BibitemOpen
  \bibfield  {author} {\bibinfo {author} {\bibfnamefont {J.~M.}\ \bibnamefont
  {Smieja}}\ and\ \bibinfo {author} {\bibfnamefont {C.~P.}\ \bibnamefont
  {Kubiak}},\ }\bibfield  {title} {\enquote {\bibinfo {title}
  {{Re(bipy-tBu)(CO)3Cl-improved catalytic activity for reduction of carbon
  dioxide: IR-spectroelectrochemical and mechanistic studies}},}\ }\href
  {https://doi.org/10.1021/ic1008363} {\bibfield  {journal} {\bibinfo
  {journal} {Inorg. Chem.}\ }\textbf {\bibinfo {volume} {49}},\ \bibinfo
  {pages} {9283--9289} (\bibinfo {year} {2010})}\BibitemShut {NoStop}%
\bibitem [{\citenamefont {Kiefer}, \citenamefont {Michocki},\ and\
  \citenamefont {Kubarych}(2021)}]{Kiefer2021}%
  \BibitemOpen
  \bibfield  {author} {\bibinfo {author} {\bibfnamefont {L.~M.}\ \bibnamefont
  {Kiefer}}, \bibinfo {author} {\bibfnamefont {L.~B.}\ \bibnamefont
  {Michocki}},\ and\ \bibinfo {author} {\bibfnamefont {K.~J.}\ \bibnamefont
  {Kubarych}},\ }\bibfield  {title} {\enquote {\bibinfo {title} {{Transmission
  Mode 2D-IR Spectroelectrochemistry of in Situ Electrocatalytic
  Intermediates}},}\ }\href {https://doi.org/10.1021/acs.jpclett.1c00504}
  {\bibfield  {journal} {\bibinfo  {journal} {J. Phys. Chem. Lett.}\ }\textbf
  {\bibinfo {volume} {12}},\ \bibinfo {pages} {3712--3717} (\bibinfo {year}
  {2021})}\BibitemShut {NoStop}%
\bibitem [{\citenamefont {Bredenbeck}, \citenamefont {Helbing},\ and\
  \citenamefont {Hamm}(2004)}]{Bredenbeck2004}%
  \BibitemOpen
  \bibfield  {author} {\bibinfo {author} {\bibfnamefont {J.}~\bibnamefont
  {Bredenbeck}}, \bibinfo {author} {\bibfnamefont {J.}~\bibnamefont
  {Helbing}},\ and\ \bibinfo {author} {\bibfnamefont {P.}~\bibnamefont
  {Hamm}},\ }\bibfield  {title} {\enquote {\bibinfo {title} {{Continuous
  scanning from picoseconds to microseconds in time resolved linear and
  nonlinear spectroscopy}},}\ }\href {https://doi.org/10.1063/1.1793891}
  {\bibfield  {journal} {\bibinfo  {journal} {Rev. Sci. Instrum.}\ }\textbf
  {\bibinfo {volume} {75}},\ \bibinfo {pages} {4462--4466} (\bibinfo {year}
  {2004})}\BibitemShut {NoStop}%
\bibitem [{\citenamefont {Yu}\ \emph {et~al.}(2005)\citenamefont {Yu},
  \citenamefont {Ye}, \citenamefont {Ionascu}, \citenamefont {Cao},\ and\
  \citenamefont {Champion}}]{yu05b}%
  \BibitemOpen
  \bibfield  {author} {\bibinfo {author} {\bibfnamefont {A.}~\bibnamefont
  {Yu}}, \bibinfo {author} {\bibfnamefont {X.}~\bibnamefont {Ye}}, \bibinfo
  {author} {\bibfnamefont {D.}~\bibnamefont {Ionascu}}, \bibinfo {author}
  {\bibfnamefont {W.}~\bibnamefont {Cao}},\ and\ \bibinfo {author}
  {\bibfnamefont {P.~M.}\ \bibnamefont {Champion}},\ }\bibfield  {title}
  {\enquote {\bibinfo {title} {{Two-color pump-probe laser spectroscopy
  instrument with picosecond time-resolved electronic delay and extended scan
  range}},}\ }\href@noop {} {\bibfield  {journal} {\bibinfo  {journal} {Rev.
  Sci. Instrum.}\ }\textbf {\bibinfo {volume} {76}},\ \bibinfo {pages} {114301}
  (\bibinfo {year} {2005})}\BibitemShut {NoStop}%
\bibitem [{\citenamefont {{Van Wilderen}}, \citenamefont {Blankenburg},\ and\
  \citenamefont {Bredenbeck}(2022)}]{VanWilderen2022}%
  \BibitemOpen
  \bibfield  {author} {\bibinfo {author} {\bibfnamefont {L.~J.}\ \bibnamefont
  {{Van Wilderen}}}, \bibinfo {author} {\bibfnamefont {L.}~\bibnamefont
  {Blankenburg}},\ and\ \bibinfo {author} {\bibfnamefont {J.}~\bibnamefont
  {Bredenbeck}},\ }\bibfield  {title} {\enquote {\bibinfo {title}
  {{Femtosecond-to-millisecond mid-IR spectroscopy of photoactive yellow
  protein uncovers structural micro-transitions of the chromophore's
  protonation mechanism}},}\ }\href {https://doi.org/10.1063/5.0091918}
  {\bibfield  {journal} {\bibinfo  {journal} {J. Chem. Phys.}\ }\textbf
  {\bibinfo {volume} {156}},\ \bibinfo {pages} {205103} (\bibinfo {year}
  {2022})}\BibitemShut {NoStop}%
\bibitem [{\citenamefont {Xin}\ \emph {et~al.}(2018)\citenamefont {Xin},
  \citenamefont {Şafak}, \citenamefont {Peng}, \citenamefont {Callahan},
  \citenamefont {Kalaydzhyan}, \citenamefont {Wang}, \citenamefont {Shtyrkova},
  \citenamefont {Zhang}, \citenamefont {Chia}, \citenamefont {Jones},
  \citenamefont {Hawthorne}, \citenamefont {Battle}, \citenamefont
  {M{\"{u}}cke}, \citenamefont {Roberts},\ and\ \citenamefont
  {K{\"{a}}rtner}}]{Xin2018}%
  \BibitemOpen
  \bibfield  {author} {\bibinfo {author} {\bibfnamefont {M.}~\bibnamefont
  {Xin}}, \bibinfo {author} {\bibfnamefont {K.}~\bibnamefont {Şafak}},
  \bibinfo {author} {\bibfnamefont {M.~Y.}\ \bibnamefont {Peng}}, \bibinfo
  {author} {\bibfnamefont {P.~T.}\ \bibnamefont {Callahan}}, \bibinfo {author}
  {\bibfnamefont {A.}~\bibnamefont {Kalaydzhyan}}, \bibinfo {author}
  {\bibfnamefont {W.}~\bibnamefont {Wang}}, \bibinfo {author} {\bibfnamefont
  {K.}~\bibnamefont {Shtyrkova}}, \bibinfo {author} {\bibfnamefont
  {Q.}~\bibnamefont {Zhang}}, \bibinfo {author} {\bibfnamefont {S.~H.}\
  \bibnamefont {Chia}}, \bibinfo {author} {\bibfnamefont {B.}~\bibnamefont
  {Jones}}, \bibinfo {author} {\bibfnamefont {T.}~\bibnamefont {Hawthorne}},
  \bibinfo {author} {\bibfnamefont {P.}~\bibnamefont {Battle}}, \bibinfo
  {author} {\bibfnamefont {O.~D.}\ \bibnamefont {M{\"{u}}cke}}, \bibinfo
  {author} {\bibfnamefont {T.}~\bibnamefont {Roberts}},\ and\ \bibinfo {author}
  {\bibfnamefont {F.~X.}\ \bibnamefont {K{\"{a}}rtner}},\ }\bibfield  {title}
  {\enquote {\bibinfo {title} {{Sub-femtosecond precision timing
  synchronization systems}},}\ }\href
  {https://doi.org/10.1016/j.nima.2017.12.040} {\bibfield  {journal} {\bibinfo
  {journal} {Nucl. Instruments Methods Phys. Res. Sect. A Accel. Spectrometers,
  Detect. Assoc. Equip.}\ }\textbf {\bibinfo {volume} {907}},\ \bibinfo {pages}
  {169--181} (\bibinfo {year} {2018})}\BibitemShut {NoStop}%
\bibitem [{\citenamefont {Konold}\ \emph {et~al.}(2020)\citenamefont {Konold},
  \citenamefont {Arik}, \citenamefont {Wei{\ss}enborn}, \citenamefont {Arents},
  \citenamefont {Hellingwerf}, \citenamefont {van Stokkum}, \citenamefont
  {Kennis},\ and\ \citenamefont {Groot}}]{Konold2020}%
  \BibitemOpen
  \bibfield  {author} {\bibinfo {author} {\bibfnamefont {P.~E.}\ \bibnamefont
  {Konold}}, \bibinfo {author} {\bibfnamefont {E.}~\bibnamefont {Arik}},
  \bibinfo {author} {\bibfnamefont {J.}~\bibnamefont {Wei{\ss}enborn}},
  \bibinfo {author} {\bibfnamefont {J.~C.}\ \bibnamefont {Arents}}, \bibinfo
  {author} {\bibfnamefont {K.~J.}\ \bibnamefont {Hellingwerf}}, \bibinfo
  {author} {\bibfnamefont {I.~H.}\ \bibnamefont {van Stokkum}}, \bibinfo
  {author} {\bibfnamefont {J.~T.}\ \bibnamefont {Kennis}},\ and\ \bibinfo
  {author} {\bibfnamefont {M.~L.}\ \bibnamefont {Groot}},\ }\bibfield  {title}
  {\enquote {\bibinfo {title} {{Confinement in crystal lattice alters entire
  photocycle pathway of the Photoactive Yellow Protein}},}\ }\href
  {https://doi.org/10.1038/s41467-020-18065-9} {\bibfield  {journal} {\bibinfo
  {journal} {Nat. Commun.}\ }\textbf {\bibinfo {volume} {11}},\ \bibinfo
  {pages} {4248} (\bibinfo {year} {2020})}\BibitemShut {NoStop}%
\bibitem [{\citenamefont {Song}\ \emph {et~al.}(2019)\citenamefont {Song},
  \citenamefont {Konar}, \citenamefont {Sechrist}, \citenamefont {Roy},
  \citenamefont {Duan}, \citenamefont {Dziurgot}, \citenamefont {Policht},
  \citenamefont {Matutes}, \citenamefont {Kubarych},\ and\ \citenamefont
  {Ogilvie}}]{Song2019}%
  \BibitemOpen
  \bibfield  {author} {\bibinfo {author} {\bibfnamefont {Y.}~\bibnamefont
  {Song}}, \bibinfo {author} {\bibfnamefont {A.}~\bibnamefont {Konar}},
  \bibinfo {author} {\bibfnamefont {R.}~\bibnamefont {Sechrist}}, \bibinfo
  {author} {\bibfnamefont {V.~P.}\ \bibnamefont {Roy}}, \bibinfo {author}
  {\bibfnamefont {R.}~\bibnamefont {Duan}}, \bibinfo {author} {\bibfnamefont
  {J.}~\bibnamefont {Dziurgot}}, \bibinfo {author} {\bibfnamefont
  {V.}~\bibnamefont {Policht}}, \bibinfo {author} {\bibfnamefont {Y.~A.}\
  \bibnamefont {Matutes}}, \bibinfo {author} {\bibfnamefont {K.~J.}\
  \bibnamefont {Kubarych}},\ and\ \bibinfo {author} {\bibfnamefont {J.~P.}\
  \bibnamefont {Ogilvie}},\ }\bibfield  {title} {\enquote {\bibinfo {title}
  {{Multispectral multidimensional spectrometer spanning the ultraviolet to the
  mid-infrared}},}\ }\href {https://doi.org/10.1063/1.5055244} {\bibfield
  {journal} {\bibinfo  {journal} {Rev. Sci. Instrum.}\ }\textbf {\bibinfo
  {volume} {90}},\ \bibinfo {pages} {1--11} (\bibinfo {year}
  {2019})}\BibitemShut {NoStop}%
\bibitem [{\citenamefont {Greetham}\ \emph {et~al.}(2012)\citenamefont
  {Greetham}, \citenamefont {Sole}, \citenamefont {Clark}, \citenamefont
  {Parker}, \citenamefont {Pollard},\ and\ \citenamefont
  {Towrie}}]{Greetham2012}%
  \BibitemOpen
  \bibfield  {author} {\bibinfo {author} {\bibfnamefont {G.~M.}\ \bibnamefont
  {Greetham}}, \bibinfo {author} {\bibfnamefont {D.}~\bibnamefont {Sole}},
  \bibinfo {author} {\bibfnamefont {I.~P.}\ \bibnamefont {Clark}}, \bibinfo
  {author} {\bibfnamefont {A.~W.}\ \bibnamefont {Parker}}, \bibinfo {author}
  {\bibfnamefont {M.~R.}\ \bibnamefont {Pollard}},\ and\ \bibinfo {author}
  {\bibfnamefont {M.}~\bibnamefont {Towrie}},\ }\bibfield  {title} {\enquote
  {\bibinfo {title} {{Time-resolved multiple probe spectroscopy}},}\ }\href
  {https://doi.org/10.1063/1.4758999} {\bibfield  {journal} {\bibinfo
  {journal} {Rev. Sci. Instrum.}\ }\textbf {\bibinfo {volume} {83}},\ \bibinfo
  {pages} {103107} (\bibinfo {year} {2012})}\BibitemShut {NoStop}%
\bibitem [{\citenamefont {Greetham}\ \emph {et~al.}(2016)\citenamefont
  {Greetham}, \citenamefont {Donaldson}, \citenamefont {Nation}, \citenamefont
  {Sazanovich}, \citenamefont {Clark}, \citenamefont {Shaw}, \citenamefont
  {Parker},\ and\ \citenamefont {Towrie}}]{Greetham2016}%
  \BibitemOpen
  \bibfield  {author} {\bibinfo {author} {\bibfnamefont {G.~M.}\ \bibnamefont
  {Greetham}}, \bibinfo {author} {\bibfnamefont {P.~M.}\ \bibnamefont
  {Donaldson}}, \bibinfo {author} {\bibfnamefont {C.}~\bibnamefont {Nation}},
  \bibinfo {author} {\bibfnamefont {I.~V.}\ \bibnamefont {Sazanovich}},
  \bibinfo {author} {\bibfnamefont {I.~P.}\ \bibnamefont {Clark}}, \bibinfo
  {author} {\bibfnamefont {D.~J.}\ \bibnamefont {Shaw}}, \bibinfo {author}
  {\bibfnamefont {A.~W.}\ \bibnamefont {Parker}},\ and\ \bibinfo {author}
  {\bibfnamefont {M.}~\bibnamefont {Towrie}},\ }\bibfield  {title} {\enquote
  {\bibinfo {title} {{A 100 kHz time-resolved multiple-probe femtosecond to
  second infrared absorption spectrometer}},}\ }\href
  {https://doi.org/10.1177/0003702816631302} {\bibfield  {journal} {\bibinfo
  {journal} {Appl. Spectrosc.}\ }\textbf {\bibinfo {volume} {70}},\ \bibinfo
  {pages} {645--653} (\bibinfo {year} {2016})}\BibitemShut {NoStop}%
\bibitem [{\citenamefont {Hammarback}\ \emph {et~al.}(2021)\citenamefont
  {Hammarback}, \citenamefont {Aucott}, \citenamefont {Bray}, \citenamefont
  {Clark}, \citenamefont {Towrie}, \citenamefont {Robinson}, \citenamefont
  {Fairlamb},\ and\ \citenamefont {Lynam}}]{Hammarback2021}%
  \BibitemOpen
  \bibfield  {author} {\bibinfo {author} {\bibfnamefont {L.~A.}\ \bibnamefont
  {Hammarback}}, \bibinfo {author} {\bibfnamefont {B.~J.}\ \bibnamefont
  {Aucott}}, \bibinfo {author} {\bibfnamefont {J.~T.}\ \bibnamefont {Bray}},
  \bibinfo {author} {\bibfnamefont {I.~P.}\ \bibnamefont {Clark}}, \bibinfo
  {author} {\bibfnamefont {M.}~\bibnamefont {Towrie}}, \bibinfo {author}
  {\bibfnamefont {A.}~\bibnamefont {Robinson}}, \bibinfo {author}
  {\bibfnamefont {I.~J.}\ \bibnamefont {Fairlamb}},\ and\ \bibinfo {author}
  {\bibfnamefont {J.~M.}\ \bibnamefont {Lynam}},\ }\bibfield  {title} {\enquote
  {\bibinfo {title} {{Direct Observation of the Microscopic Reverse of the
  Ubiquitous Concerted Metalation Deprotonation Step in C-H Bond Activation
  Catalysis}},}\ }\href {https://doi.org/10.1021/jacs.0c10409} {\bibfield
  {journal} {\bibinfo  {journal} {J. Am. Chem. Soc.}\ }\textbf {\bibinfo
  {volume} {143}},\ \bibinfo {pages} {1356--1364} (\bibinfo {year}
  {2021})}\BibitemShut {NoStop}%
\bibitem [{\citenamefont {Laptenok}\ \emph {et~al.}(2018)\citenamefont
  {Laptenok}, \citenamefont {Gil}, \citenamefont {Hall}, \citenamefont
  {Lukacs}, \citenamefont {Iuliano}, \citenamefont {Jones}, \citenamefont
  {Greetham}, \citenamefont {Donaldson}, \citenamefont {Miyawaki},
  \citenamefont {Tonge},\ and\ \citenamefont {Meech}}]{Laptenok2018}%
  \BibitemOpen
  \bibfield  {author} {\bibinfo {author} {\bibfnamefont {S.~P.}\ \bibnamefont
  {Laptenok}}, \bibinfo {author} {\bibfnamefont {A.~A.}\ \bibnamefont {Gil}},
  \bibinfo {author} {\bibfnamefont {C.~R.}\ \bibnamefont {Hall}}, \bibinfo
  {author} {\bibfnamefont {A.}~\bibnamefont {Lukacs}}, \bibinfo {author}
  {\bibfnamefont {J.~N.}\ \bibnamefont {Iuliano}}, \bibinfo {author}
  {\bibfnamefont {G.~A.}\ \bibnamefont {Jones}}, \bibinfo {author}
  {\bibfnamefont {G.~M.}\ \bibnamefont {Greetham}}, \bibinfo {author}
  {\bibfnamefont {P.}~\bibnamefont {Donaldson}}, \bibinfo {author}
  {\bibfnamefont {A.}~\bibnamefont {Miyawaki}}, \bibinfo {author}
  {\bibfnamefont {P.~J.}\ \bibnamefont {Tonge}},\ and\ \bibinfo {author}
  {\bibfnamefont {S.~R.}\ \bibnamefont {Meech}},\ }\bibfield  {title} {\enquote
  {\bibinfo {title} {{Infrared spectroscopy reveals multi-step multi-timescale
  photoactivation in the photoconvertible protein archetype dronpa}},}\ }\href
  {https://doi.org/10.1038/s41557-018-0073-0} {\bibfield  {journal} {\bibinfo
  {journal} {Nat. Chem.}\ }\textbf {\bibinfo {volume} {10}},\ \bibinfo {pages}
  {845--852} (\bibinfo {year} {2018})}\BibitemShut {NoStop}%
\bibitem [{\citenamefont {Koyama}, \citenamefont {Donaldson},\ and\
  \citenamefont {Orr-Ewing}(2017)}]{Koyama2017}%
  \BibitemOpen
  \bibfield  {author} {\bibinfo {author} {\bibfnamefont {D.}~\bibnamefont
  {Koyama}}, \bibinfo {author} {\bibfnamefont {P.~M.}\ \bibnamefont
  {Donaldson}},\ and\ \bibinfo {author} {\bibfnamefont {A.~J.}\ \bibnamefont
  {Orr-Ewing}},\ }\bibfield  {title} {\enquote {\bibinfo {title} {{Femtosecond
  to microsecond observation of the photochemical reaction of
  1,2-di(quinolin-2-yl)disulfide with methyl methacrylate}},}\ }\href
  {https://doi.org/10.1039/c7cp01784g} {\bibfield  {journal} {\bibinfo
  {journal} {Phys. Chem. Chem. Phys.}\ }\textbf {\bibinfo {volume} {19}},\
  \bibinfo {pages} {12981--12991} (\bibinfo {year} {2017})}\BibitemShut
  {NoStop}%
\bibitem [{\citenamefont {Bartels}\ \emph {et~al.}(2007)\citenamefont
  {Bartels}, \citenamefont {Cerna}, \citenamefont {Kistner}, \citenamefont
  {Thoma}, \citenamefont {Hudert}, \citenamefont {Janke},\ and\ \citenamefont
  {Dekorsy}}]{Bartels2007}%
  \BibitemOpen
  \bibfield  {author} {\bibinfo {author} {\bibfnamefont {A.}~\bibnamefont
  {Bartels}}, \bibinfo {author} {\bibfnamefont {R.}~\bibnamefont {Cerna}},
  \bibinfo {author} {\bibfnamefont {C.}~\bibnamefont {Kistner}}, \bibinfo
  {author} {\bibfnamefont {A.}~\bibnamefont {Thoma}}, \bibinfo {author}
  {\bibfnamefont {F.}~\bibnamefont {Hudert}}, \bibinfo {author} {\bibfnamefont
  {C.}~\bibnamefont {Janke}},\ and\ \bibinfo {author} {\bibfnamefont
  {T.}~\bibnamefont {Dekorsy}},\ }\bibfield  {title} {\enquote {\bibinfo
  {title} {{Ultrafast time-domain spectroscopy based on high-speed asynchronous
  optical sampling}},}\ }\href {https://doi.org/10.1063/1.2714048} {\bibfield
  {journal} {\bibinfo  {journal} {Rev. Sci. Instrum.}\ }\textbf {\bibinfo
  {volume} {78}},\ \bibinfo {pages} {035107} (\bibinfo {year}
  {2007})}\BibitemShut {NoStop}%
\bibitem [{\citenamefont {Antonucci}\ \emph {et~al.}(2012)\citenamefont
  {Antonucci}, \citenamefont {Solinas}, \citenamefont {Bonvalet},\ and\
  \citenamefont {Joffre}}]{Antonucci2012}%
  \BibitemOpen
  \bibfield  {author} {\bibinfo {author} {\bibfnamefont {L.}~\bibnamefont
  {Antonucci}}, \bibinfo {author} {\bibfnamefont {X.}~\bibnamefont {Solinas}},
  \bibinfo {author} {\bibfnamefont {A.}~\bibnamefont {Bonvalet}},\ and\
  \bibinfo {author} {\bibfnamefont {M.}~\bibnamefont {Joffre}},\ }\bibfield
  {title} {\enquote {\bibinfo {title} {{Asynchronous optical sampling with
  arbitrary detuning between laser repetition rates}},}\ }\href
  {https://doi.org/10.1364/oe.20.017928} {\bibfield  {journal} {\bibinfo
  {journal} {Opt. Express}\ }\textbf {\bibinfo {volume} {20}},\ \bibinfo
  {pages} {17928----17937} (\bibinfo {year} {2012})}\BibitemShut {NoStop}%
\bibitem [{\citenamefont {Antonucci}\ \emph {et~al.}(2015)\citenamefont
  {Antonucci}, \citenamefont {Bonvalet}, \citenamefont {Solinas}, \citenamefont
  {Daniault},\ and\ \citenamefont {Joffre}}]{Antonucci2015}%
  \BibitemOpen
  \bibfield  {author} {\bibinfo {author} {\bibfnamefont {L.}~\bibnamefont
  {Antonucci}}, \bibinfo {author} {\bibfnamefont {A.}~\bibnamefont {Bonvalet}},
  \bibinfo {author} {\bibfnamefont {X.}~\bibnamefont {Solinas}}, \bibinfo
  {author} {\bibfnamefont {L.}~\bibnamefont {Daniault}},\ and\ \bibinfo
  {author} {\bibfnamefont {M.}~\bibnamefont {Joffre}},\ }\bibfield  {title}
  {\enquote {\bibinfo {title} {{Arbitrary-detuning asynchronous optical
  sampling with amplified laser systems}},}\ }\href
  {https://doi.org/10.1364/oe.23.027931} {\bibfield  {journal} {\bibinfo
  {journal} {Opt. Express}\ }\textbf {\bibinfo {volume} {23}},\ \bibinfo
  {pages} {27931--27940} (\bibinfo {year} {2015})}\BibitemShut {NoStop}%
\bibitem [{\citenamefont {Solinas}\ \emph {et~al.}(2017)\citenamefont
  {Solinas}, \citenamefont {Antonucci}, \citenamefont {Bonvalet},\ and\
  \citenamefont {Joffre}}]{Solinas2017}%
  \BibitemOpen
  \bibfield  {author} {\bibinfo {author} {\bibfnamefont {X.}~\bibnamefont
  {Solinas}}, \bibinfo {author} {\bibfnamefont {L.}~\bibnamefont {Antonucci}},
  \bibinfo {author} {\bibfnamefont {A.}~\bibnamefont {Bonvalet}},\ and\
  \bibinfo {author} {\bibfnamefont {M.}~\bibnamefont {Joffre}},\ }\bibfield
  {title} {\enquote {\bibinfo {title} {{Multiscale control and rapid scanning
  of time delays ranging from picosecond to millisecond}},}\ }\href
  {https://doi.org/10.1364/oe.25.017811} {\bibfield  {journal} {\bibinfo
  {journal} {Opt. Express}\ }\textbf {\bibinfo {volume} {25}},\ \bibinfo
  {pages} {17811} (\bibinfo {year} {2017})}\BibitemShut {NoStop}%
\bibitem [{\citenamefont {Antonucci}\ \emph {et~al.}(2020)\citenamefont
  {Antonucci}, \citenamefont {Solinas}, \citenamefont {Bonvalet},\ and\
  \citenamefont {Joffre}}]{Antonucci2020}%
  \BibitemOpen
  \bibfield  {author} {\bibinfo {author} {\bibfnamefont {L.}~\bibnamefont
  {Antonucci}}, \bibinfo {author} {\bibfnamefont {X.}~\bibnamefont {Solinas}},
  \bibinfo {author} {\bibfnamefont {A.}~\bibnamefont {Bonvalet}},\ and\
  \bibinfo {author} {\bibfnamefont {M.}~\bibnamefont {Joffre}},\ }\bibfield
  {title} {\enquote {\bibinfo {title} {{Electronic measurement of femtosecond
  time delays for arbitrary-detuning asynchronous optical sampling}},}\ }\href
  {https://doi.org/10.1364/oe.393887} {\bibfield  {journal} {\bibinfo
  {journal} {Opt. Express}\ }\textbf {\bibinfo {volume} {28}},\ \bibinfo
  {pages} {18251} (\bibinfo {year} {2020})}\BibitemShut {NoStop}%
\bibitem [{\citenamefont {Fl{\"{o}}ry}\ \emph {et~al.}(2023)\citenamefont
  {Fl{\"{o}}ry}, \citenamefont {Stummer}, \citenamefont {Pupeikis},
  \citenamefont {Willenberg}, \citenamefont {Nussbaum-Lapping}, \citenamefont
  {Kaksis}, \citenamefont {Camargo}, \citenamefont {Barkauskas}, \citenamefont
  {Phillips}, \citenamefont {Keller}, \citenamefont {G}, \citenamefont
  {Pugzlys},\ and\ \citenamefont {Baltu{\v{s}}ka}}]{Floery2023}%
  \BibitemOpen
  \bibfield  {author} {\bibinfo {author} {\bibfnamefont {T.}~\bibnamefont
  {Fl{\"{o}}ry}}, \bibinfo {author} {\bibfnamefont {V.}~\bibnamefont
  {Stummer}}, \bibinfo {author} {\bibfnamefont {J.}~\bibnamefont {Pupeikis}},
  \bibinfo {author} {\bibfnamefont {B.}~\bibnamefont {Willenberg}}, \bibinfo
  {author} {\bibfnamefont {A.}~\bibnamefont {Nussbaum-Lapping}}, \bibinfo
  {author} {\bibfnamefont {E.}~\bibnamefont {Kaksis}}, \bibinfo {author}
  {\bibfnamefont {F.~V.~A.}\ \bibnamefont {Camargo}}, \bibinfo {author}
  {\bibfnamefont {M.}~\bibnamefont {Barkauskas}}, \bibinfo {author}
  {\bibfnamefont {C.~R.}\ \bibnamefont {Phillips}}, \bibinfo {author}
  {\bibfnamefont {U.}~\bibnamefont {Keller}}, \bibinfo {author} {\bibfnamefont
  {C.}~\bibnamefont {G}}, \bibinfo {author} {\bibfnamefont {A.}~\bibnamefont
  {Pugzlys}},\ and\ \bibinfo {author} {\bibfnamefont {A.}~\bibnamefont
  {Baltu{\v{s}}ka}},\ }\bibfield  {title} {\enquote {\bibinfo {title}
  {{Rapid-scan nonlienar time-resolved spectroscopy over arbitrary delay
  intervals.}}}\ }\href@noop {} {\bibfield  {journal} {\bibinfo  {journal}
  {Ultrafast Sci.}\ }\textbf {\bibinfo {volume} {3}},\ \bibinfo {pages} {27}
  (\bibinfo {year} {2023})}\BibitemShut {NoStop}%
\bibitem [{\citenamefont {Harmer}, \citenamefont {Linz},\ and\ \citenamefont
  {Gabbe}(1969)}]{Harmer1969}%
  \BibitemOpen
  \bibfield  {author} {\bibinfo {author} {\bibfnamefont {A.~L.}\ \bibnamefont
  {Harmer}}, \bibinfo {author} {\bibfnamefont {A.}~\bibnamefont {Linz}},\ and\
  \bibinfo {author} {\bibfnamefont {D.~R.}\ \bibnamefont {Gabbe}},\ }\bibfield
  {title} {\enquote {\bibinfo {title} {{Fluorescence of Nd3+ in lithium yttrium
  fluoride}},}\ }\href@noop {} {\bibfield  {journal} {\bibinfo  {journal} {J.
  Phys. Chem. Solids}\ }\textbf {\bibinfo {volume} {30}},\ \bibinfo {pages}
  {1483--1491} (\bibinfo {year} {1969})}\BibitemShut {NoStop}%
\bibitem [{\citenamefont {Best}(2007)}]{Best2007}%
  \BibitemOpen
  \bibfield  {author} {\bibinfo {author} {\bibfnamefont {R.~E.}\ \bibnamefont
  {Best}},\ }\href@noop {} {\emph {\bibinfo {title} {{Phase-Locked Loops
  Design, Simulation, and Applications}}}},\ \bibinfo {edition} {6th}\ ed.\
  (\bibinfo  {publisher} {McGraw-Hill Education},\ \bibinfo {address} {New
  York},\ \bibinfo {year} {2007})\BibitemShut {NoStop}%
\bibitem [{\citenamefont {Donaldson}\ \emph {et~al.}(2023)\citenamefont
  {Donaldson}, \citenamefont {Greetham}, \citenamefont {Middleton},
  \citenamefont {Luther}, \citenamefont {Zanni}, \citenamefont {Hamm},\ and\
  \citenamefont {Krummel}}]{Donaldson2023}%
  \BibitemOpen
  \bibfield  {author} {\bibinfo {author} {\bibfnamefont {P.~M.}\ \bibnamefont
  {Donaldson}}, \bibinfo {author} {\bibfnamefont {G.~M.}\ \bibnamefont
  {Greetham}}, \bibinfo {author} {\bibfnamefont {C.~T.}\ \bibnamefont
  {Middleton}}, \bibinfo {author} {\bibfnamefont {B.~M.}\ \bibnamefont
  {Luther}}, \bibinfo {author} {\bibfnamefont {M.}~\bibnamefont {Zanni}},
  \bibinfo {author} {\bibfnamefont {P.}~\bibnamefont {Hamm}},\ and\ \bibinfo
  {author} {\bibfnamefont {A.~T.}\ \bibnamefont {Krummel}},\ }\bibfield
  {title} {\enquote {\bibinfo {title} {{Breaking Barriers in Ultrafast
  Spectroscopy and Imaging Using 100 kHz Amplified Yb-Laser Systems}},}\
  }\href@noop {} {\bibfield  {journal} {\bibinfo  {journal} {Acc. Chem. Res.}\
  ,\ \bibinfo {pages} {accepted for publication, arXiv:2303.04250}} (\bibinfo
  {year} {2023})}\BibitemShut {NoStop}%
\bibitem [{\citenamefont {Hamm}, \citenamefont {Kaindl},\ and\ \citenamefont
  {Stenger}(2000)}]{ham00b}%
  \BibitemOpen
  \bibfield  {author} {\bibinfo {author} {\bibfnamefont {P.}~\bibnamefont
  {Hamm}}, \bibinfo {author} {\bibfnamefont {R.~A.}\ \bibnamefont {Kaindl}},\
  and\ \bibinfo {author} {\bibfnamefont {J.}~\bibnamefont {Stenger}},\
  }\bibfield  {title} {\enquote {\bibinfo {title} {{Noise suppression in
  femtosecond mid-infrared light sources}},}\ }\href@noop {} {\bibfield
  {journal} {\bibinfo  {journal} {Opt. Lett.}\ }\textbf {\bibinfo {volume}
  {25}},\ \bibinfo {pages} {1798--1800} (\bibinfo {year} {2000})}\BibitemShut
  {NoStop}%
\bibitem [{\citenamefont {Farrell}\ \emph {et~al.}(2020)\citenamefont
  {Farrell}, \citenamefont {Ostrander}, \citenamefont {Jones}, \citenamefont
  {Yakami}, \citenamefont {Dicke}, \citenamefont {Middleton}, \citenamefont
  {Hamm},\ and\ \citenamefont {Zanni}}]{Farrell2020}%
  \BibitemOpen
  \bibfield  {author} {\bibinfo {author} {\bibfnamefont {K.~M.}\ \bibnamefont
  {Farrell}}, \bibinfo {author} {\bibfnamefont {J.~S.}\ \bibnamefont
  {Ostrander}}, \bibinfo {author} {\bibfnamefont {A.~C.}\ \bibnamefont
  {Jones}}, \bibinfo {author} {\bibfnamefont {B.~R.}\ \bibnamefont {Yakami}},
  \bibinfo {author} {\bibfnamefont {S.~S.}\ \bibnamefont {Dicke}}, \bibinfo
  {author} {\bibfnamefont {C.~T.}\ \bibnamefont {Middleton}}, \bibinfo {author}
  {\bibfnamefont {P.}~\bibnamefont {Hamm}},\ and\ \bibinfo {author}
  {\bibfnamefont {M.~T.}\ \bibnamefont {Zanni}},\ }\bibfield  {title} {\enquote
  {\bibinfo {title} {{Shot-to-shot 2D IR spectroscopy at 100 kHz using a Yb
  laser and custom-designed electronics}},}\ }\href@noop {} {\bibfield
  {journal} {\bibinfo  {journal} {Opt. Express}\ }\textbf {\bibinfo {volume}
  {28}},\ \bibinfo {pages} {33584} (\bibinfo {year} {2020})}\BibitemShut
  {NoStop}%
\bibitem [{\citenamefont {Asbury}, \citenamefont {Wang},\ and\ \citenamefont
  {Lian}(2002)}]{asbury02}%
  \BibitemOpen
  \bibfield  {author} {\bibinfo {author} {\bibfnamefont {J.~B.}\ \bibnamefont
  {Asbury}}, \bibinfo {author} {\bibfnamefont {Y.~Q.}\ \bibnamefont {Wang}},\
  and\ \bibinfo {author} {\bibfnamefont {T.~Q.}\ \bibnamefont {Lian}},\
  }\bibfield  {title} {\enquote {\bibinfo {title} {{Time-dependent vibration
  Stokes shift during solvation: Experiment and theory}},}\ }\href@noop {}
  {\bibfield  {journal} {\bibinfo  {journal} {Bull. Chem. Soc. Jpn.}\ }\textbf
  {\bibinfo {volume} {75}},\ \bibinfo {pages} {973--983} (\bibinfo {year}
  {2002})}\BibitemShut {NoStop}%
\bibitem [{\citenamefont {Uhmann}\ \emph {et~al.}(1991)\citenamefont {Uhmann},
  \citenamefont {Becker}, \citenamefont {Taran},\ and\ \citenamefont
  {Siebert}}]{uhmann91}%
  \BibitemOpen
  \bibfield  {author} {\bibinfo {author} {\bibfnamefont {W.}~\bibnamefont
  {Uhmann}}, \bibinfo {author} {\bibfnamefont {A.}~\bibnamefont {Becker}},
  \bibinfo {author} {\bibfnamefont {C.}~\bibnamefont {Taran}},\ and\ \bibinfo
  {author} {\bibfnamefont {F.}~\bibnamefont {Siebert}},\ }\bibfield  {title}
  {\enquote {\bibinfo {title} {{Time-resolved FT-IR absorption-spectroscopy
  using a step-scan interferometer}},}\ }\href@noop {} {\bibfield  {journal}
  {\bibinfo  {journal} {Appl. Spectrosocpy}\ }\textbf {\bibinfo {volume}
  {45}},\ \bibinfo {pages} {390--397} (\bibinfo {year} {1991})}\BibitemShut
  {NoStop}%
\bibitem [{\citenamefont {Gerwert}(1993)}]{Gerwert1993}%
  \BibitemOpen
  \bibfield  {author} {\bibinfo {author} {\bibfnamefont {K.}~\bibnamefont
  {Gerwert}},\ }\bibfield  {title} {\enquote {\bibinfo {title} {{Molecular
  Reaction-Mechanisms of Proteins As Monitored by Time-Resolved Ftir
  Spectroscopy}},}\ }\href {https://doi.org/10.1515/BC.1999.115} {\bibfield
  {journal} {\bibinfo  {journal} {Curr. Opin. Struct. Biol.}\ }\textbf
  {\bibinfo {volume} {3}},\ \bibinfo {pages} {769--773} (\bibinfo {year}
  {1993})}\BibitemShut {NoStop}%
\bibitem [{\citenamefont {Kottke}, \citenamefont {L{\'{o}}renz-Fonfr{\'{i}}a},\
  and\ \citenamefont {Heberle}(2017)}]{Kottke2017}%
  \BibitemOpen
  \bibfield  {author} {\bibinfo {author} {\bibfnamefont {T.}~\bibnamefont
  {Kottke}}, \bibinfo {author} {\bibfnamefont {V.~A.}\ \bibnamefont
  {L{\'{o}}renz-Fonfr{\'{i}}a}},\ and\ \bibinfo {author} {\bibfnamefont
  {J.}~\bibnamefont {Heberle}},\ }\bibfield  {title} {\enquote {\bibinfo
  {title} {{The Grateful Infrared: Sequential Protein Structural Changes
  Resolved by Infrared Difference Spectroscopy}},}\ }\href
  {https://doi.org/10.1021/acs.jpcb.6b09222} {\bibfield  {journal} {\bibinfo
  {journal} {J. Phys. Chem. B}\ }\textbf {\bibinfo {volume} {121}},\ \bibinfo
  {pages} {335--350} (\bibinfo {year} {2017})}\BibitemShut {NoStop}%
\bibitem [{\citenamefont {Klocke}\ \emph {et~al.}(2018)\citenamefont {Klocke},
  \citenamefont {Mangold}, \citenamefont {Allmendinger}, \citenamefont {Hugi},
  \citenamefont {Geiser}, \citenamefont {Jouy}, \citenamefont {Faist},\ and\
  \citenamefont {Kottke}}]{Klocke2018}%
  \BibitemOpen
  \bibfield  {author} {\bibinfo {author} {\bibfnamefont {J.~L.}\ \bibnamefont
  {Klocke}}, \bibinfo {author} {\bibfnamefont {M.}~\bibnamefont {Mangold}},
  \bibinfo {author} {\bibfnamefont {P.}~\bibnamefont {Allmendinger}}, \bibinfo
  {author} {\bibfnamefont {A.}~\bibnamefont {Hugi}}, \bibinfo {author}
  {\bibfnamefont {M.}~\bibnamefont {Geiser}}, \bibinfo {author} {\bibfnamefont
  {P.}~\bibnamefont {Jouy}}, \bibinfo {author} {\bibfnamefont {J.}~\bibnamefont
  {Faist}},\ and\ \bibinfo {author} {\bibfnamefont {T.}~\bibnamefont
  {Kottke}},\ }\bibfield  {title} {\enquote {\bibinfo {title} {{Single-Shot
  Sub-microsecond Mid-infrared Spectroscopy on Protein Reactions with Quantum
  Cascade Laser Frequency Combs}},}\ }\href
  {https://doi.org/10.1021/acs.analchem.8b02531} {\bibfield  {journal}
  {\bibinfo  {journal} {Anal. Chem.}\ }\textbf {\bibinfo {volume} {90}},\
  \bibinfo {pages} {10494--10500} (\bibinfo {year} {2018})}\BibitemShut
  {NoStop}%
\bibitem [{\citenamefont {Stritt}, \citenamefont {Jawurek},\ and\ \citenamefont
  {Hauser}(2020)}]{Stritt2020}%
  \BibitemOpen
  \bibfield  {author} {\bibinfo {author} {\bibfnamefont {P.}~\bibnamefont
  {Stritt}}, \bibinfo {author} {\bibfnamefont {M.}~\bibnamefont {Jawurek}},\
  and\ \bibinfo {author} {\bibfnamefont {K.}~\bibnamefont {Hauser}},\
  }\bibfield  {title} {\enquote {\bibinfo {title} {{Application of tunable
  quantum cascade lasers to monitor dynamics of bacteriorhodopsin in the mid-IR
  spectral range}},}\ }\href {https://doi.org/10.3233/bsi-200195} {\bibfield
  {journal} {\bibinfo  {journal} {Biomed. Spectrosc. Imaging}\ }\textbf
  {\bibinfo {volume} {9}},\ \bibinfo {pages} {55--61} (\bibinfo {year}
  {2020})}\BibitemShut {NoStop}%
\bibitem [{\citenamefont {Schubert}\ \emph {et~al.}(2022)\citenamefont
  {Schubert}, \citenamefont {Langner}, \citenamefont {Ehrenberg}, \citenamefont
  {Lorenz-Fonfria},\ and\ \citenamefont {Heberle}}]{Schubert2022}%
  \BibitemOpen
  \bibfield  {author} {\bibinfo {author} {\bibfnamefont {L.}~\bibnamefont
  {Schubert}}, \bibinfo {author} {\bibfnamefont {P.}~\bibnamefont {Langner}},
  \bibinfo {author} {\bibfnamefont {D.}~\bibnamefont {Ehrenberg}}, \bibinfo
  {author} {\bibfnamefont {V.~A.}\ \bibnamefont {Lorenz-Fonfria}},\ and\
  \bibinfo {author} {\bibfnamefont {J.}~\bibnamefont {Heberle}},\ }\bibfield
  {title} {\enquote {\bibinfo {title} {{Protein conformational changes and
  protonation dynamics probed by a single shot using
  quantum-cascade-laser-based IR spectroscopy}},}\ }\href
  {https://doi.org/10.1063/5.0088526} {\bibfield  {journal} {\bibinfo
  {journal} {J. Chem. Phys.}\ }\textbf {\bibinfo {volume} {156}},\ \bibinfo
  {pages} {204201} (\bibinfo {year} {2022})}\BibitemShut {NoStop}%
\bibitem [{\citenamefont {Herbst}, \citenamefont {Heyne},\ and\ \citenamefont
  {Diller}(2002)}]{herbst02}%
  \BibitemOpen
  \bibfield  {author} {\bibinfo {author} {\bibfnamefont {R.}~\bibnamefont
  {Herbst}}, \bibinfo {author} {\bibfnamefont {K.}~\bibnamefont {Heyne}},\ and\
  \bibinfo {author} {\bibfnamefont {R.}~\bibnamefont {Diller}},\ }\bibfield
  {title} {\enquote {\bibinfo {title} {{Femtosecond infrared spectroscopy of
  bacteriorhodopsin chromophore isomerization}},}\ }\href@noop {} {\bibfield
  {journal} {\bibinfo  {journal} {Science}\ }\textbf {\bibinfo {volume}
  {297}},\ \bibinfo {pages} {822--825} (\bibinfo {year} {2002})}\BibitemShut
  {NoStop}%
\bibitem [{\citenamefont {Gross}\ \emph {et~al.}(2009)\citenamefont {Gross},
  \citenamefont {Schumann}, \citenamefont {Wolf}, \citenamefont {Herbst},
  \citenamefont {Diller}, \citenamefont {Friedman},\ and\ \citenamefont
  {Sheves}}]{Gross2009a}%
  \BibitemOpen
  \bibfield  {author} {\bibinfo {author} {\bibfnamefont {R.}~\bibnamefont
  {Gross}}, \bibinfo {author} {\bibfnamefont {C.}~\bibnamefont {Schumann}},
  \bibinfo {author} {\bibfnamefont {M.~M.}\ \bibnamefont {Wolf}}, \bibinfo
  {author} {\bibfnamefont {J.}~\bibnamefont {Herbst}}, \bibinfo {author}
  {\bibfnamefont {R.}~\bibnamefont {Diller}}, \bibinfo {author} {\bibfnamefont
  {N.}~\bibnamefont {Friedman}},\ and\ \bibinfo {author} {\bibfnamefont
  {M.}~\bibnamefont {Sheves}},\ }\bibfield  {title} {\enquote {\bibinfo {title}
  {{Ultrafast protein conformational alterations in bacteriorhodopsin and its
  locked analogue BR5.12}},}\ }\href {https://doi.org/10.1021/jp810042f}
  {\bibfield  {journal} {\bibinfo  {journal} {J. Phys. Chem. B}\ }\textbf
  {\bibinfo {volume} {113}},\ \bibinfo {pages} {7851--7860} (\bibinfo {year}
  {2009})}\BibitemShut {NoStop}%
\bibitem [{\citenamefont {Smith}\ \emph {et~al.}(1987)\citenamefont {Smith},
  \citenamefont {Braiman}, \citenamefont {Myers}, \citenamefont {Pardoen},
  \citenamefont {Courtin}, \citenamefont {Winkel}, \citenamefont {Lugtenburg},\
  and\ \citenamefont {Mathies}}]{smith87}%
  \BibitemOpen
  \bibfield  {author} {\bibinfo {author} {\bibfnamefont {S.~O.}\ \bibnamefont
  {Smith}}, \bibinfo {author} {\bibfnamefont {M.~S.}\ \bibnamefont {Braiman}},
  \bibinfo {author} {\bibfnamefont {A.~B.}\ \bibnamefont {Myers}}, \bibinfo
  {author} {\bibfnamefont {J.~A.}\ \bibnamefont {Pardoen}}, \bibinfo {author}
  {\bibfnamefont {J.~M.~L.}\ \bibnamefont {Courtin}}, \bibinfo {author}
  {\bibfnamefont {C.}~\bibnamefont {Winkel}}, \bibinfo {author} {\bibfnamefont
  {J.}~\bibnamefont {Lugtenburg}},\ and\ \bibinfo {author} {\bibfnamefont
  {R.~A.}\ \bibnamefont {Mathies}},\ }\bibfield  {title} {\enquote {\bibinfo
  {title} {{Vibrational analysis of the all-trans-retinal chromophore in
  light-adapted bacteriorhodopsin}},}\ }\href@noop {} {\bibfield  {journal}
  {\bibinfo  {journal} {J. Am. Chem. Soc.}\ }\textbf {\bibinfo {volume}
  {109}},\ \bibinfo {pages} {3108--3125} (\bibinfo {year} {1987})}\BibitemShut
  {NoStop}%
\bibitem [{\citenamefont {Polland}\ \emph {et~al.}(1986)\citenamefont
  {Polland}, \citenamefont {Franz}, \citenamefont {Zinth}, \citenamefont
  {Kaiser}, \citenamefont {K{\"{o}}lling},\ and\ \citenamefont
  {Oesterhelt}}]{Polland1986}%
  \BibitemOpen
  \bibfield  {author} {\bibinfo {author} {\bibfnamefont {H.~J.}\ \bibnamefont
  {Polland}}, \bibinfo {author} {\bibfnamefont {M.~A.}\ \bibnamefont {Franz}},
  \bibinfo {author} {\bibfnamefont {W.}~\bibnamefont {Zinth}}, \bibinfo
  {author} {\bibfnamefont {W.}~\bibnamefont {Kaiser}}, \bibinfo {author}
  {\bibfnamefont {E.}~\bibnamefont {K{\"{o}}lling}},\ and\ \bibinfo {author}
  {\bibfnamefont {D.}~\bibnamefont {Oesterhelt}},\ }\bibfield  {title}
  {\enquote {\bibinfo {title} {{Early Picosecond Events in the Photocycle of
  Bacteriorhodopsin}},}\ }\href {https://doi.org/10.1016/S0006-3495(86)83692-X}
  {\bibfield  {journal} {\bibinfo  {journal} {Biophys. J.}\ }\textbf {\bibinfo
  {volume} {49}},\ \bibinfo {pages} {651--662} (\bibinfo {year}
  {1986})}\BibitemShut {NoStop}%
\bibitem [{\citenamefont {Helbing}\ and\ \citenamefont
  {Bonmarin}(2009)}]{Helbing09}%
  \BibitemOpen
  \bibfield  {author} {\bibinfo {author} {\bibfnamefont {J.}~\bibnamefont
  {Helbing}}\ and\ \bibinfo {author} {\bibfnamefont {M.}~\bibnamefont
  {Bonmarin}},\ }\bibfield  {title} {\enquote {\bibinfo {title} {{Vibrational
  circular dichroism signal enhancement using self-heterodyning with
  elliptically polarized laser pulses}},}\ }\href@noop {} {\bibfield  {journal}
  {\bibinfo  {journal} {J. Chem. Phys.}\ }\textbf {\bibinfo {volume} {131}},\
  \bibinfo {pages} {174507} (\bibinfo {year} {2009})}\BibitemShut {NoStop}%
\bibitem [{\citenamefont {Hamm}(1995)}]{Hamm1995}%
  \BibitemOpen
  \bibfield  {author} {\bibinfo {author} {\bibfnamefont {P.}~\bibnamefont
  {Hamm}},\ }\bibfield  {title} {\enquote {\bibinfo {title} {{Coherent effects
  in femtosecond infrared spectroscopy}},}\ }\href@noop {} {\bibfield
  {journal} {\bibinfo  {journal} {Chem. Phys.}\ }\textbf {\bibinfo {volume}
  {200}},\ \bibinfo {pages} {415--429} (\bibinfo {year} {1995})}\BibitemShut
  {NoStop}%
\bibitem [{\citenamefont {Wynne}\ and\ \citenamefont
  {Hochstrasser}(1995)}]{Wynne1995}%
  \BibitemOpen
  \bibfield  {author} {\bibinfo {author} {\bibfnamefont {K.}~\bibnamefont
  {Wynne}}\ and\ \bibinfo {author} {\bibfnamefont {R.~M.}\ \bibnamefont
  {Hochstrasser}},\ }\bibfield  {title} {\enquote {\bibinfo {title} {{The
  theory of ultrafast vibrational spectroscopy}},}\ }\href
  {https://doi.org/10.1016/0301-0104(95)00012-D} {\bibfield  {journal}
  {\bibinfo  {journal} {Chem. Phys.}\ }\textbf {\bibinfo {volume} {193}},\
  \bibinfo {pages} {211--236} (\bibinfo {year} {1995})}\BibitemShut {NoStop}%
\end{thebibliography}

%aipnum4-2.bst 2019-01-14 (MD) hand-edited version of apsrev4-1.bst
%Control: key (0)
%Control: author (8) initials jnrlst
%Control: editor formatted (1) identically to author
%Control: production of article title (0) allowed
%Control: page (1) range
%Control: year (1) truncated
%Control: production of eprint (0) enabled
%

\end{document}